\documentclass[prb,aps,showpacs,floatfix,floats,nobalancelastpage,twocolumn]{revtex4-1}
\usepackage{graphicx}
\usepackage{graphics}
\usepackage{latexsym,amsmath,amssymb,bm,euscript}
\usepackage{color}
\usepackage{dcolumn}
\usepackage{bm}
\usepackage{epstopdf}
\usepackage{times}
\usepackage{multirow}
\usepackage{wasysym}
\usepackage{subfigure}
\usepackage{bbm}

\def\ra{\rangle}
\def\la{\langle}

\def\Hc{{\rm H.c.}}

\begin{document}

\title{Effects of short-ranged interactions on the Kane-Mele model without discrete particle-hole symmetry}
\author{Hsin-Hua Lai}
\affiliation{National High Magnetic Field Laboratory, Florida State University, Tallahassee, Florida 32310, USA}
\author{Hsiang-Hsuan Hung}
\affiliation{Department of Physics, The University of Texas at Austin, Austin, Texas 78712, USA}
\date{\today}
\pacs{}

\begin{abstract}
We study the effects of short-ranged interactions on the $Z_2$ topological insulator phase, also known as the quantum spin Hall phase, in the Kane-Mele model at half-filling with staggered potentials which explicitly breaks the discrete particle-hole symmetry. Within Hartree-Fock mean-field analysis, we conclude that the on-site repulsive interactions help stabilize the topological phase (quantum spin Hall) against the staggered potentials by enlarging the regime of the topological phase along the axis of the ratio of the staggered potential strength and the spin-orbit coupling. In sharp contrast, the on-site attractive interactions destabilize the topological phase. We also examine the attractive interaction case by means of the unbiased determinant projector quantum Monte Carlo and the results are qualitatively consistent with the Hartree-Fock picture.
\end{abstract}
\maketitle

\section{Introduction}\label{Introduction}
Topological insulators (TI) \cite{fu2007,moore2007, moore2007,roy2009,hasan2010,qi2011}   are perhaps some of the most intriguing states of matter found in recent years. Much interest in these is initially motivated by the first experimental realization in HgTe/(Hg,Cd)Te quantum wells which shows quantized quantum spin Hall conductance protected by the time-reversal symmetry (TRS). \cite{Pankratov1987, Bernevig2006, Kšnig2007} The most important of all is that the existence of the topological state and many of its properties can be well understood in some noninteracting models. Among them, the $Z_2$ topological insulator (TI) or quantum spin Hall (QSH) insulator can be realized in the noninteracting Kane-Mele (KM) model \cite{kane2005a}.
The KM model can be described as two copies of the original Haldane model on the two-dimensional (2D) honeycomb lattice \cite{haldane1988} with the usual real nearest-neighbor hopping and imaginary second-neighbor hopping which arises from the spin-orbit couplings (SOC). Due to the simplicity of the KM model, it has recently served as the theoretical framework to study the phase diagram of the interacting $Z_2$ TIs. \cite{rachel2010, yu2011,zheng2011,hohenadler2011,Budich:prb12,wuwei2012,hohenadler2012,Griset2012,Hung2013,hung2013b,lang2013} Based on the numerical studies on the Kane-Mele-Hubbard (KMH) model, several exotic quantum phases have been proposed, including the quantum spin liquid (QSL). \cite{hohenadler2012, Zong2013} Even though several phase diagrams have been proposed in the KMH model, so far there have just a few studies focusing on the interacting effects on the stabilization of the QSH. Specifically, the question about if the short-ranged interactions enlarge (stabilize) the QSH regime found in the noninteracting KM model has not been paid much attention in most of the previous studies.

Recently, Ref.~\onlinecite{Hung2013} studies the KMH model with third-neighbor hoppings at half-filling using determinant projector quantum Monte Carlo (QMC) and conclude that the short-ranged on-site repulsive Hubbard interactions tend to push the QSH phase boundary to larger threshold values of the third-neighbor hoppings resulting in the stabilization of the QSH in a larger parameter regime due to the interactions. The generalized KM model in Ref.~\onlinecite{Hung2013} as well as the original KM model at half-filling possesses the discrete particle-hole symmetry (PHS), \cite{zheng2011} $\{c_{j \alpha} \rightarrow \pm c^\dagger_{j\alpha}, c^\dagger_{j \alpha} \rightarrow \pm c_{j\alpha}\}$ with $\pm$ for $j\in \{A,B\}$, where $A$ and $B$ are the sublattice labeling. The natural question is if the stabilization of QSH against other effective gap closing perturbations due to short-ranged interactions can exist in a KMH model with lower symmetries. To answer this question, at least in some simple model, in this work we focus on the KMH model {\it without} PHS at half-filling. We explicitly break the PHS by including the staggered potentials in the original KM model.\cite{kane2005a} It is well known that in the noninteracting limit, the model contains both QSH and trivial phase depending on the ratio of the strength of the staggered potentials  ($m_a$) and that of spin-orbital couplings ($\lambda_{so}$). The phase transition between these two phases happens when the band gaps close. In the presence of the short-ranged interactions, it is expected that the band gaps get renormalized, which causes the shift of the phase boundary. Within Hartree-Fock mean-field approach,\cite{Cai2008} we find that the short-ranged repulsive interactions increase the threshold value of $m_a/\lambda_{so}$ to widen the QSH regime, suggesting in this model the QSH state is more stabilized against the staggered potentials by the short-ranged repulsive interactions. On the contrary, the short-ranged attractive interactions destabilize the QSH, making it more fragile to the staggered potentials. 

In order to go beyond the Hartree-Fock mean-field picture, numerical tools are highly demanding. The determinant projector QMC can be used to detect the topological phase but, however, in the case with repulsive interactions it suffers from the well-known sign problem. On the other hand, the QMC can still be performed on the attractive interaction side. Within the Hartree-Fock picture, opposite to the result in the repulsive interaction case, the on-site attractive interactions tend to shrink the QSH regime. This picture is confirmed by the QMC analysis in Sec.~\ref{Sec:QMC}, which may imply that our Hartree-Fock picture in the repulsive interaction side is possibly plausible. In appendix~\ref{App:RG_critical} we also perform renormalization group (RG) analysis at the tree-level in the gapless critical phase at the phase boundary between the topologically trivial and nontrivial phases where the band gaps close to form Dirac points. We conclude that such a coarse-grained picture {\it can not} correctly predict a shift of the phase boundary.

This paper is organized as follows. In Sec.~\ref{Sec:model} we define explicitly the model Hamiltonian that we will study. In Sec.~\ref{Subsec:Hartree-Fock} we introduce the Hartree-Fock approach to decouple the short-ranged interactions and in Sec.~\ref{Subsec:Ncalc} we numerically solve for the transition point between the QSH and the trivial phase by finding the gap closing point self-consistently. In Sec.~\ref{Sec:QMC} we present the QMC results in the KMH model with attractive interactions. In Sec.~\ref{Sec:Discussion} we conclude with some discussions. In Appendix~\ref{App:RG_critical} we show, at the long-wavelength description, the tree-level RG analysis on the critical phase at band gap closing point. In Appendix~\ref{App:UV} we provide the Hartree-Fock mean-field studies on the case with the more extended interactions, including both on-site interaction $U$ and the nearest-neighbor $V$.
\section{Kane-Mele-Hubbard model with staggered on-site potentials}\label{Sec:model}
\begin{figure}[t]
   \centering
   \includegraphics[width=1.5 in]{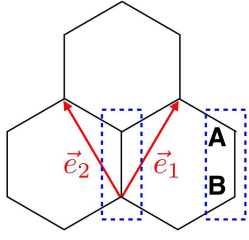}
   \caption{Illustration of the honeycomb lattice. There are two sublattices per unit cell labeled as $A$ and $B$. $\vec{e}_{1/2} = (\pm 1/2, \sqrt{3}/2)$ are the two vectors connecting the same sublattices in different unit cells. We set the lattice constant to be 1.}
   \label{Fig:honeycomb}
\end{figure}
The model we are focusing in this work is the honeycomb Kane-Mele-Hubbard model in the presence of staggered potentials at half-filling \cite{kane2005a}. The honeycomb lattice is shown in Fig.~\ref{Fig:honeycomb} and the model Hamiltonian is $H = H_{KMs} + H_{U}$, where
\begin{eqnarray}
\nonumber && H_{KMs}= t \sum_{\la jk \ra} \sum_{\sigma} c^\dagger_{j \sigma} c_{k \sigma} + i \lambda_{so} \sum_{\la \la jk \ra \ra } \sum_{\sigma} \sigma c^\dagger_{j \sigma} \nu_{jk} c_{k\sigma} \\
&& \hspace{4cm} + \sum_{j} \sum_{\sigma}  \epsilon_j  m_j c^\dagger_{j \sigma} c_{j \sigma}, \\
&& H_{U} = U \sum_{j}  n_{j\uparrow} n_{j\downarrow},\label{Eq:Hubbard}
\end{eqnarray}
with $\epsilon_j = \pm 1$ for sublattice $j \in \{ A, B\}$. The $\nu_{jk} = +1(-1)$ for (counter-)clockwise second-neighbor hopping, and $U$ is the strength of the interaction that can be positive and negative. The positive $U$ corresponds to the on-site Hubbard repulsion. We theoretically consider the negative $U$ side not only for the completeness of this work but also for comparing the results obtained in the mean-field analysis in Sec.~\ref{Subsec:Ncalc} with those obtained in the unbiased determinant projector QMC studies in Sec.~\ref{Sec:QMC}. The QMC works in the negative $U$ side but suffers from the sign problem in the positive $U$ side. Without the staggered potentials, the finite spin-orbit couplings break the spin SU$(2)$ down to U$(1)$ and also break the sublattice symmetry, while the TRS and PHS still remain. The presence of the staggered potentials further break the PHS.

For the case of the $U>0$, the previous numerical studies \cite{hohenadler2012,yu2011, zheng2011, hohenadler2011,Budich:prb12,wuwei2012} on the KMH model shows that the magnetic transition to the topologically trivial insulator phase such as antiferro-magnetic Mott insulator (AFM) happens at a quite large $U/t$ value. The critical value of the Hubbard repulsion for $\lambda_{so}/t=0.1$ is roughly $U_c/t \simeq 5$, \cite{hohenadler2012} and the critical $U_c/t$ is even larger for larger $\lambda_{so}/t$. On the other hand, for $U<0$, the theoretical mean-field studies by Yuan {\it et. al} \cite{Yuan2012} suggests that the critical value of the attractive Hubbard interaction is $U_c/t < - 2$ for finite spin-orbit couplings. If $U$ is smaller than the critical value, there would be a possible phase transition to the superconducting state whose bulk is still insulating but superconductivity appears near the edges. \cite{Yuan2012} Suggested by the previous numerical studies in both $U$-axis (negative or positive), in this work we restrict our analysis within $|U/t| < 2$ to ignore possibly magnetic phase transitions and within the regime it is appropriate to ignore the magnetic phases. 

For clarification, from now on we replace the site labeling $j$ with $j = \{{\bf r}, a\}$, where ${\bf r}$ runs over the Bravais lattice of unit cells of the honeycomb network and $a$ runs over the two sites ($A$ and $B$) in the unit cell. We define the potential on sublattice $A (B)$ as $m_A(m_B)$. In the noninteracting limit, the Hamiltonian in the momentum space is $H_{KMs} = \sum_{k\in {\bf B. Z.}} \Psi^\dagger_{\bf k} h_{KMs}({\bf k}) \Psi_{\bf k}$, with
\begin{eqnarray}
\nonumber && h_{KMs}({\bf k})= \begin{pmatrix}
m_A  & t f({\bf k}) \\
t f^*({\bf k}) & - m_B
\end{pmatrix}
\otimes \mathbbm{1}_{2\times 2} + \\
&& \hspace{3cm} +
\begin{pmatrix}
2 \lambda_{so} g({\bf k}) & 0 \\
0& - 2 \lambda_{so} g({\bf k})
\end{pmatrix}\otimes \sigma_z.~~\label{Eq:h_KMs}
\end{eqnarray}
$\sigma_z$ is the Pauli matrix and $ \Psi^{T}_{\bf k} \equiv \begin{pmatrix}
c_{\uparrow}({\bf k},A) & c_{\uparrow} ({\bf k},B) & c_{\downarrow} ({\bf k}, A) & c_{\downarrow} ({\bf k}, B)
\end{pmatrix}$. $f({\bf k}) \equiv 1 + e^{i {\bf k} \cdot \vec{e}_1} + e^{i {\bf k} \cdot \vec{e}_2}$, and $g({\bf k}) \equiv \sin ({\bf k}\cdot \vec{e}_1) - \sin({\bf k}\cdot \vec{e}_2) - \sin ({\bf k} \cdot (\vec{e}_1 - \vec{e}_2)).$ Without lack of generality, we choose $m_A$, $m_B$, and $\lambda_{so}$ to be positive.
\subsection{Hartree-Fock approach}\label{Subsec:Hartree-Fock}
In order to take the on-site interaction into considerations, we use Hartree-Fock mean-field decoupling approach to decouple it as
\begin{eqnarray}
\nonumber H_U & = & U \sum_{({\bf r},a)} n_{\uparrow} ( {\bf r},a) n_{\downarrow}({\bf r},a) \\
  & \simeq & \frac{U}{2} \sum_{({\bf r},a)} \bigg{[} \left\la n( {\bf r},a) \right\ra n({\bf r},a) + \left\la s_z ({\bf r},a) \right\ra s_z ({\bf r},a) \bigg{]},~~ \label{Eq: Hubbard in Hartree-Fock}
 \end{eqnarray}
with $n\equiv n_\uparrow + n_\downarrow$, and $s_z \equiv n_\uparrow - n_\downarrow$.
We have explicitly neglected the constant $\la n_j \ra^2$ appearing in the Hartree-Fock decoupling since they only shift the total energy. The terms $\la c^\dagger_\sigma c_{\bar{\sigma}}\ra $ also vanish since they do not conserve $S^z$. Since there is no local magnetic field at each site, the local magnetization is zero, which means the second term in the Eq.~(\ref{Eq: Hubbard in Hartree-Fock}) vanishes. The on-site interaction within the Hartree-Fock picture adds density modulations to the diagonal elements in the KMs matrix (\ref{Eq:h_KMs}). In addition, due to the translational symmetry $\la n({\bf r},a) \ra= \la n (a)\ra \equiv \la n_a \ra$ and the crucial point is that the densities at site $A$ and $B$ are different due to the presence of the staggered potential. In the momentum space, $h({\bf k}) = h_{KMs}({\bf k}) + h_u ({\bf k})$, with  $h_{KMs}$ defined in Eq.~(\ref{Eq:h_KMs}) and
 \begin{eqnarray}
 h_u ({\bf k}) = \frac{U}{2}\begin{pmatrix}
 \la n_A \ra \\
 0 &  \la n_B \ra
 \end{pmatrix}
 \otimes \mathbbm{1}_{2\times2}.
 \end{eqnarray}

Since the phase boundary between the QSH and the trivial phase in the KM model is located at the parameter regime in which the band gaps close, the key point is to find when the band gaps close in the presence of the short-ranged interactions. Within Hartree-Fock picture, since no symmetry is spontaneous broken, the band gaps associated with the full Hamiltonian matrix still close at ${\bf K} = - {\bf K'} = \{ 4\pi/3,0\}$ in the Brillouin zone ({\bf B.Z.}). Near ${\bf K}$ and ${\bf K'}$, the off-diagonal elements in Eq.~(\ref{Eq:h_KMs}) become linear in ${\bf k}$ and vanish exactly at ${\bf K}$ and ${\bf K'}$. Since the two points are related by the TRS, the gaps must simultaneously close at both ${\bf K}$  and ${\bf K'}$. It is then sufficient to examine the gap near ${\bf K}$ only. Focusing on ${\bf K}$, we find that the two of the four bands with eigenvalues $E_1 = m_A - 2\lambda_{so} g(K) + \frac{U}{2} \la n_A \ra$ and $E_2 = - m_B + 2 \lambda_{so} g(K) + \frac{U}{2} \la n_B \ra$ would get inverted by tuning the mass $m_{A/B}$ and $\lambda_{so}$. We then define the gap function $\Delta({\bf K})$ as
 \begin{eqnarray}
 \nonumber \Delta({\bf K}) = m_A + m_B -4 \lambda_{so} g({\bf K}) + \frac{U}{2} \bigg{(} \la n_A \ra - \la n_B \ra \bigg{)},\\ \label{Eq:gap_function}
 \end{eqnarray}
with $g(\pm {\bf K}) =\pm 3\sqrt{3}/2$. When $\Delta({\bf K}) >0$, the bands are not inverted and we are in the topologically trivial phase. When $\Delta({\bf K}) <0$, the bands are inverted and we are in the topological phase. The transition happens at $\Delta( {\bf K}) = 0$ when the band gaps close to form Dirac points.

\subsection{Self-consistent numerical calculations}\label{Subsec:Ncalc}
For simplicity, in the self-consistent numerical calculations, we set $m_A = m_B\equiv \lambda_m$. Before jumping to the numerical calculations, we first give a simple physical picture here. Since the presence of the staggered potentials lowers the chemical potential at sublattice $B$, the total number density at sublattice $B$ is expected to be higher than that in sublattice $A$. The contribution from the last term in Eq.~(\ref{Eq:gap_function}) is then negative (positive) for repulsive (attractive) interaction. For fixed $\lambda_{so}$, the gap function reaches zero at larger (smaller) mass compared with the case without the on-site repulsive (attractive) interaction. Therefore, the regime of the topological phase is {\it enlarged} ({\it shrunk}) in the presence of the on-site repulsive (attractive) interaction.

\begin{figure}[t]
\subfigure[]
{\label{Fig:onsite_positive} \includegraphics[width=1.5 in]{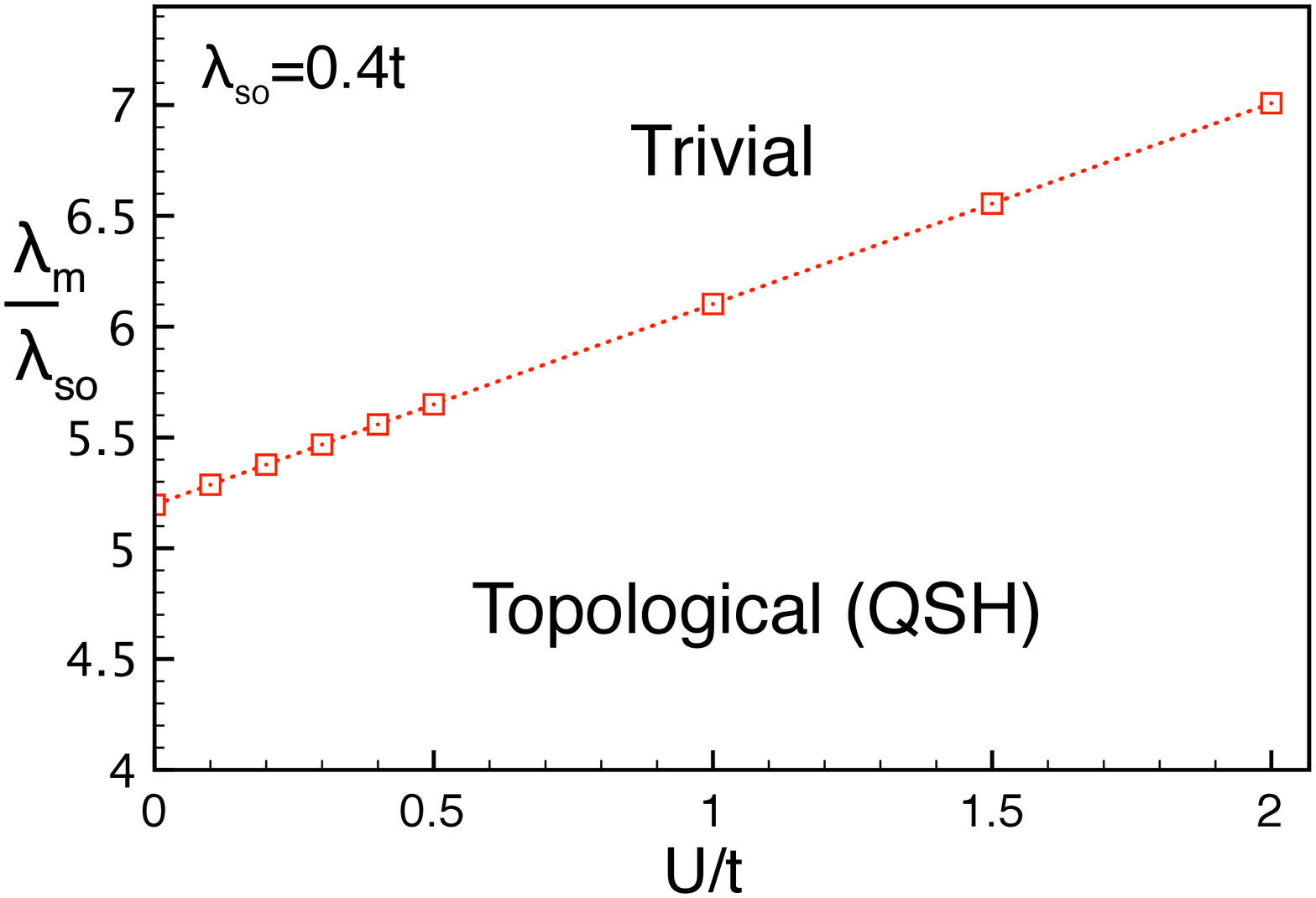}}
\subfigure[]
{\label{Fig:onsite_negative} \includegraphics[width = 1.5 in]{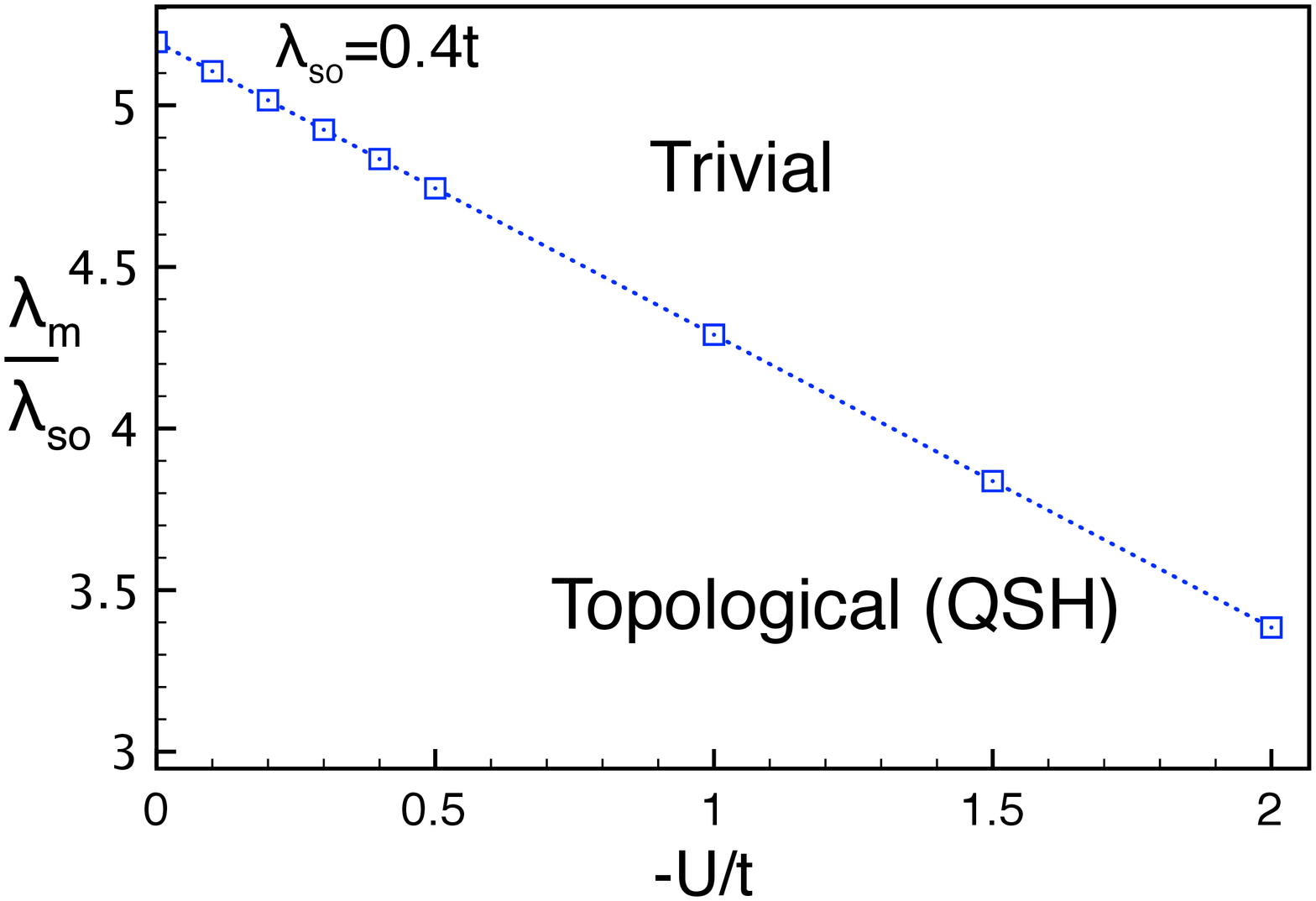}}
\caption{The phase boundary between the topologically nontrivial phase (QSH) and the trivial phase within the Hartree-Fock picture. (a) The numerical data in the repulsive interaction case, during the numerics, we set $t=1$, $\lambda_{so} = 0.4$ and $U>0$. The Hubbard repulsion tends to screen the stagger potentials to push the boundary of the nontrivial phase toward the trivial side, which widens the regime of the nontrivial phase. (b) The numerical date in the attractive interaction case, $U=-|U|$. Contrary to the result in (a), the attractive interactions enhances the staggered potential and shrink the regime of the topological phase.}
\label{Fig:onsite_U}
\end{figure}

In the numerical calculations, the honeycomb lattice consists of $200\times200$ unit cells and we choose to set $t =1$  and $\lambda_{so} =0.4$. The result is shown in Fig.~\ref{Fig:onsite_U}. We can see that the on-site repulsive interaction {\it widens} the regime of QSH, Fig.~\ref{Fig:onsite_positive}. The physical picture is that the on-site Hubbard repulsion screens the staggered potentials resulting in stabilizing the topological phase against the staggered potentials. The attractive $U$ on the other hand enhances the staggered potentials, which leads to the opposite result shown in Fig.~\ref{Fig:onsite_negative}. In order to have a more complete and unbiased analysis, next section we perform QMC to compare qualitatively with the results obtained in the Hartree-Fock picture here. For the repulsive interaction case, there is a sign problem and can not be accessed by the QMC. We thus focus on the attractive interaction case in which the QMC analysis is free of sign-problem in Sec.~\ref{Sec:QMC}
\section{Sign-free determinant projector QMC studies in the Kane-Mele model with on-site attractive interaction}\label{Sec:QMC}
In this section, we will study the KM model with staggered
potentials using the unbiased projector QMC method. Due to a sign
problem in the QMC in the case of the repulsive Hubbard interaction,
Eq.~(\ref{Eq:Hubbard}) with $U>0$, we will focus on the case with
the attractive interaction and compare the result qualitatively with
that in the Hartree-Fock picture. We remark that the presence of the
staggered potentials breaks the PHS, so even {\it at half-filling},
the QMC result in the attractive interaction case is {\it different}
from that in the repulsive case. We will see that the behavior given
by the QMC is consistent with that from the Hartree-Fock analysis,
where the attractive interactions destabilize the QSH phase.

In the QMC method, the expectation value of an arbitrary observable
$\hat{O}$ is evaluated as
\begin{eqnarray}
\langle \hat{O}\rangle =\lim_{\Theta \to \infty}\frac{\langle \psi_T
|e^{- \frac{\Theta }{2}H} \hat{O} e^{- \frac{\Theta
}{2}H}|\psi_T\rangle}{\langle \psi_T |e^{-\Theta H}| \psi_T\rangle},
\label{eqn:expection}
\end{eqnarray}
where $\Theta$ is the projective parameter. The trivial wave
function $|\psi_T\rangle$ is required to have nonvanishing overlap
with the ground state wave function $|\psi_0\rangle$, i.e. $\langle
\psi_T |\psi_0\rangle \ne 0$. Numerically, the projection operator
$e^{-\Theta H}$ is discretized into $e^{-\Delta \tau H}$,
 written as $e^{-\Theta H}=(e^{-\Delta \tau H})^M$, where $\Theta=\Delta \tau M$; $M$ is the number of
time slices and $\Delta \tau$ is chosen as a small number. In the
first-order Suzuki-Trotter decomposition, we can have
\begin{eqnarray}
e^{-\Delta \tau H} \simeq e^{-\Delta \tau H_{KMs}}e^{-\Delta \tau
H_U}, \label{eqn:suzuki}
\end{eqnarray}
where $H_{KMs}$ is the Hamiltonian of the KM model with staggered
potentials. $H_U$ is the attractive Hubbard on-site interaction
\begin{eqnarray} H_U=-|U|\sum_j
n_{j\uparrow}n_{j\downarrow},
\end{eqnarray}
where $n_{j,\sigma}=c^{\dag}_{j,\sigma}c_{j,\sigma}$,
$n_j=\sum_{\sigma}c^{\dag}_{j,\sigma}c_{j,\sigma}$. At half-filling,
we can recast $H_U$ as
\begin{eqnarray}
H_U=-\frac{|U|}{2}\sum_{j} (n_j-1)^2,
\end{eqnarray}
For the attractive interaction case, we can implement the exact
Hubbard-Stratonovich transformation\cite{ hirsch1985}
\begin{eqnarray}
e^{-\Delta \tau H_U}=e^{\Delta \tau
\frac{|U|}{2}(n_j-1)^2}=\frac{1}{2}\sum_{s=\pm
1}e^{s\alpha(n_j-1)},\label{eqn:HStransformation}
\end{eqnarray}
where $\alpha=\textrm{arcCosh}(e^{\frac{|U|\Delta \tau}{2}})$. By
the implementation, the denominator of Eq. (\ref{eqn:expection}),
 also named the projector partition function,\cite{sorella1989} can be numerically evaluated
as \cite{assaad2002,hohenadler2012,zheng2011}
\begin{eqnarray}
 & & \langle \psi_T | e^{-\Theta H} |\psi_T\rangle \nonumber \\ & \cong &\langle
\psi_T | \prod^M_{\tau=1} e^{-\Delta \tau H_{KMs}}e^{-\Delta \tau
H_{U,\tau}} |\psi_T\rangle\nonumber\\
&\sim &\sum_{\lbrace s_{i,\tau} \rbrace} \prod_{\sigma} \textrm{Tr}
\Big(\prod^M_{\tau=1} e^{-\Delta \tau
\sum_{i,j}c^{\dag}_{i,\sigma}[{\bf
H^{\sigma}_{KMs}}]_{ij}c_{j,\sigma}}e^{\alpha
s_{i,\tau}(n_{i,\sigma}-\frac{1}{2})}
\Big)\nonumber\\
&\sim &\sum_{\lbrace s_{i,\tau} \rbrace} \Bigg\{ \det
\Big(O_{\uparrow}[s_{i,\tau}] \Big)\det
\Big(O_{\downarrow}[s_{i,\tau}] \Big) \Bigg\} \nonumber \\
&=& \sum_{\lbrace s_{i,\tau} \rbrace} w[s_{i,\tau}],
\label{eqn:paritionfun}
\end{eqnarray}
where $\det \Big( O_{\sigma} [s_{i,\tau}] \Big)$ is the matrix
determinant of $e^{-\Delta \tau H_{KMs}}e^{-\Delta \tau H_U}$ with
fermion trace $"\textrm{Tr}"$ and at a given auxiliary configuration
$\lbrace s_{i,\tau}\rbrace$. In the QSH with the attractive
interaction, we have $O_{\uparrow}[s_{i,\tau}] = \Big(
O_{\downarrow}[s_{i,\tau}] \Big)^*$. Therefore, the weight
$w[s_{i,\tau}]>0$ and the sampling in Eq. (\ref{eqn:paritionfun}) is
sign-free. In the current literature, $\Delta \tau t=0.05$ and
$\Theta t=40$ are used through the content. The eigenstates of the
noninteracting Hamiltonian $H_{KMs}$ is used as the trial wave
function.

To characterize the QSH state and a trivial insulator, we usually
need to evaluate the $Z_2$ topological invariant.\cite{ fu2007} In
the KM model, however, the existence of the staggered potential
breaks the inversion symmetry; thus the conventional approach to
evaluate the $Z_2$ index, $\Delta$,
\begin{eqnarray}
(-1)^{\Delta}=\prod_{{\mathbf k_i} \in TRIM} \tilde{\eta}_{i}
\end{eqnarray}
is not applicable. However, the model still does not break the $S_z$
conservation, and thus the spin Chern number $C_{\sigma}$ is a
proper topological measurement to probe the phase transition. The
spin Chern number formalism has been proposed using the QMC method
to acquire the zero-frequency single-particle Green's functions
$G(i\omega=0,\vec{k})$ \cite{WangPRB,Wang_PRX, Wang2013, hung2013b, meng2013} and then
construct projector operators in terms of the R-zero eigenstates of
the Green's functions to evaluate the spin Chern number\cite{
avron1983}
\begin{align}
C_\sigma = \frac{i}{2\pi}\int d^2k\;
\epsilon^{\mu\nu}\text{Tr}\left[ P_\sigma(k)\;\partial_{\mu}
P_\sigma(k) \;\partial_{\nu}P_\sigma(k) \right],
\label{eqn:c1P}\end{align}  where $P_\sigma(k)=\sum_n
|v_{nk\sigma}\rangle \langle v_{nk\sigma}|$ is the single particle
spectral projector onto R-zero modes; i.e.,
$G(0,\vec{k})|v_{nk\sigma}\rangle=\lambda_{nk\sigma}|v_{nk\sigma}\rangle$
and $\lambda_{nk\sigma}>0$. Although the spin Chern number has been
shown to suffer strong finite-size effect, it is still useful to
determine the topological phase boundary in the interaction case by
observing the dramatic change in the topological number.\cite{WangPRB, 
Wang_PRX, Wang2013,
hung2013b, meng2013}

\begin{figure}[t!]
\centering
\includegraphics[width=2.5in]{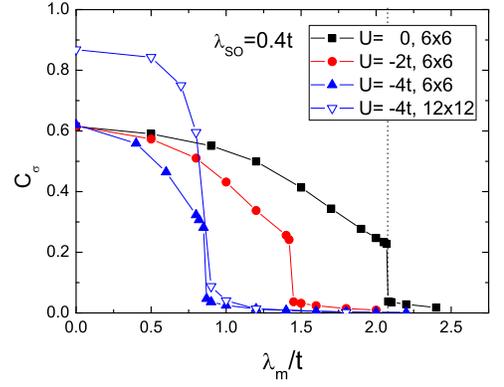}
\caption{The spin Chern number $C_{\sigma}$ vs the magnitudes of
staggered potentials $\lambda_{m}/t$ using QMC at
$\lambda_{SO}=0.4t$. The dot line indicates the noninteracting phase
boundary $\lambda^c_m=3\sqrt{3}\lambda_{SO}=2.0784t$. The solid
symbols denote the $6 \times 6$ QMC results at $U=0,-2t$ and $-4t$.
The hollow symbols denote the $12 \times 12$ QMC at $U=-4t$.
  }
   \label{Fig:QMC}
\end{figure}

The QMC results are shown in Fig. \ref{Fig:QMC}. The spin-orbital
coupling is used at $\lambda_{SO}=0.4t$. In the noninteracting
limit, the topological phase boundary is identified at
$\lambda^c_m=3\sqrt{3}\lambda_{SO}=2.0784t$,\cite{ haldane1988,
kane2005b} depicted as the dot line in Fig. \ref{Fig:QMC}. As
$\lambda_m < \lambda^c_m$, the system is a QSH (left hand side of
the dot line); otherwise it is a trivial insulator.
It is obvious to see that $C_{\sigma}$ has strong finite-size
effect, since it is poorly quantized. However, even on the
noninteracting $6\times 6$ cluster at the critical point, a
significant drop is clearly seen and the location is consistent with
the analytical prediction. Upon introducing the attractive
interaction, the topological phase boundary shifts away from
$\lambda^c_m$. One can observe that at $U=-|U|=-2t$ and $-4t$, the
critical points can be roughly identified at $1.4t$ and $0.9t$ by
QMC simulation. Upon increasing the strength of the Hubbard
interaction, the critical point moves towards to the QSH phase, i.e.
the attractive interaction destabilizes the QSH phase. For the case
of $U/t = -2$, we can see that the boundary is roughly at
$\lambda^c_m/\lambda_{so} \simeq 1.4/0.4 = 3.5$, which is quite
close to the Hartree-Fock analysis focusing on  the ${\bf K}$ point
in {\bf B.Z.} in
Sec.~\ref{Subsec:Hartree-Fock}$,\lambda^c_m/\lambda_{so} \simeq
3.4$.

On the other hand, the location of the boundary is subject to weaker
finite-size effect. For $U=-4t$, the locations where the spin Chern
numbers drop for $6\times 6$ and $12 \times 12$ clusters are fairly
close. As a consequence, one can merely perform the QMC simulation
on a small-size cluster to determine the topological phase boundary.
Moreover, in comparison with the $6\times 6$ and $12 \times 12$
clusters, the computed spin Chern numbers show closer to the
quantized number with increasing system sizes, so the poor
quantization is expected to vanish in the thermodynamic limit.
Therefore, by the unbiased QMC simulation, we can arrive at a
summary that the attractive interaction brings a minus effect to the
mass, which is consistent with the Hartree-Fock approach.

\section{Discussion}\label{Sec:Discussion}
In this work, we study the effects of the short-ranged interaction on the Kane-Mele model with staggered potentials which breaks the discrete particle-hole symmetry. Within the Hartree-Fock mean-field approach, we conclude that the short-ranged repulsive interactions help stabilize the topological phase (QSH) against the staggered potentials by enlarging the regime of the topological phase (QSH) along the axis of the ratio of the staggered potential strength and the spin-orbit coupling. In sharp contrast, the short-ranged attractive interactions destabilize the topological phase (QSH), making it more fragile to the staggered potentials. Beyond the Hartree-Fock picture, we study the interaction effects using the projector QMC. Due to the sign problem, the KM model with repulsive interaction can not be accessed by the QMC, and thus we focus on the case of the attractive interaction. The QMC results show that the attractive interactions shrink the QSH regime, which is consistent with that in the Hartree-Fock approach. The qualitative consistency between these two approaches in the case of attractive interaction may imply that the Hartree-Fock analysis in the repulsive side is plausible. Because QMC suffers from sign problem in this case, other numerical tool such as Dynamical Mean Field Theory (DMFT)\cite{Maier2005,Kotliar2006,Weiwu2010,wuwei2012, Budich2013} may be used to confirm our conjecture in the future work.

The boundary shifts due to the presence of the on-site interaction in this model can not be explained by the continuum low-energy theory at the critical phase in which the gaps close to form Dirac points. Explicitly, in Appendix~\ref{App:RG_critical} we focus on the low-energy descriptions at the critical phase and perform tree-level RG analysis. The straightforward thoughts is that even though the local four-fermion interactions are irrelevant in the critical phase, before they flow to negligible values under RG they can still generate a finite bilinear mass term, which can possibly shift the phase boundary. However, we find that the tree-level RG corrections completely cancel each other, which gives no generation of a bilinear mass term. Hence we conclude that the boundary shift can only be captured by the physics of the lattice model and can not be captured by the coarse-grained continuum theory around ${\bf K}$ and ${\bf K'}$.

In this work, we only consider the on-site interaction effects on the Kane-Mele model with staggered potentials. It is interesting to consider a more extended interaction case such as a $U$-$V$ model (including both on-site $U$ and nearest-neighbor $V$). In Appendix~\ref{App:UV} we check the case with both on-site repulsive (attractive) Hubbard $U$ and a nearest neighbor repulsive (attractive) $V$. The inclusion of the nearest-neighbor $V$ complicates the analysis and we find that whether or not the topological phase is stabilized due to the short-ranged interactions depends on the details of the competition between the on-site $U$ and the nearest-neighbor $V$. From the low energy analysis focusing on ${\bf K}$ point in the {\bf B.Z.}, we conclude that within the Hartree-Fock picture if $U$ is dominant over $V$ ($U > 6V$ ) the qualitative result obtained from the case with only on-site $U$, repulsive (attractive) interactions stabilize (destabilize) the topological phase, is still correct in the $U$-$V$ model. However, from the studies of the Kane-Mele-$U$-$V$ model, it may suggest a long-ranged repulsive interaction such as Coulomb interaction that decays very slow may completely destabilize the QSH phase.

As a final remark, there is a recent paper addressing the Hubbard interaction effects on the topological insulators properties using the slave-rotor formalism.\cite{Miguel2013} In that approach, both the spin orbital coupling, $\lambda_{so}$, and the staggered potential, $\lambda_m$ are {\it renormalized} and the situation of whether or not the Hubbard interactions stabilize the QSH phase is not clear yet, depending on the details of the renormalized ratios of $\lambda^*_{so}/\lambda^*_m$. Though, it is very likely that the results obtained in that approach are consistent with what we find in this work. Besides, the slave-rotor formalism \cite{Florens2004} can also be applied to the case in the generalized Kane-Mele model with third-neighbor hoppings\cite{hung2013b} in which the Hartree-Fock mean-field approach fails to predict any boundary shift due to the presence of Hubbard interaction found in the QMC studies.

\acknowledgments
One of the authors, H.-H. Lai, would like to thank Kun Yang for helpful discussions. H.-H. Lai acknowledges the support by the National Science Foundation through grant No. DMR-1004545.  H.-H. Hung acknowledges computational support from the Center for Scientific Computing at the CNSI and MRL: an NSF MRSEC (DMR-1121053) and NSF CNS-0960316, as well as the support from ARO Grant No. W911NF-09-1-0527, and Grant No.W911NF-12-1-0573 from the Army Research Office with funding from the DARPA OLE Program.
\appendix

\section{RG analysis of the critical phase in the Kane-Mele model with weak $U$ in the presence of staggered potentials}\label{App:RG_critical}
In this model, since $S^z$ is still conserved, the spin-up and spin-down Hamiltonian can be treated separately. For each spin species, we can diagonalize the Hamiltonian matrix for spectra. There are $2$ bands for each spin species. The bands can be characterized by the eigenvector-eigenenergy pairs $\{ \vec{v}^\alpha_{b}({\bf k}),\epsilon^\alpha_{b} ({\bf k})\}$, where $b=1,2$ are band indices. The Hamiltonian can be diagonalized by rewriting the original fermion fields in terms of the complex fermion fields $d^\alpha_b({\bf k})$ in the diagonal basis,
\begin{eqnarray}
c_{\alpha}({\bf r},a) = \sqrt{\frac{1}{N_{uc}}}\sum_{b=1,2}\sum_{{\bf k}\in{\bf B.~Z.}}v^\alpha_{b}({\bf k},a) d^\alpha_b ({\bf k}) e^{i {\bf k}\cdot {\bf r}},~~
\end{eqnarray}
where $N_{uc}$ is the number of unit cells and the complex fermion field $f$ satisfies the usual anti commutation relation $ \{ d^{\alpha \dagger}_{b}({\bf k}), d^{\alpha'}_{b' }({\bf k'}) \} = \delta_{\alpha \alpha'}\delta_{b b'} \delta_{{\bf k} {\bf k'}}$. In terms of the new complex fermion fields, the Hamiltonian becomes
\begin{eqnarray}
H_{KMs} = \sum_{b=1,2} \sum_{\alpha=\uparrow,\downarrow} \sum_{{\bf k}\in{\bf B.~Z.}}\epsilon^\alpha_{b}({\bf k}) d^{\alpha\dagger}_b({\bf k}) d^\alpha_b ({\bf k}).
\end{eqnarray}
At the critical phase $(\lambda^c_m = 3\sqrt{3} \lambda_{so})$, the gaps close at momentums ${\bf K}$ and ${\bf K'} = - {\bf K}$. Around these points, only the spin-down fermions are gapless at ${\bf K}$ and spin-up fermions are gapless at ${\bf K'}$. As far as the long-wavelength (low-energy) description is concerned, we can focus on the ${\bf K}$ and ${\bf K'}$ points and perform expansion around theses points by introducing a small momentum shift $\delta{\bf k}$.

For the low-energy description at momentum ${\bf K}$, we find that only the spin-down fermions are gapless and can expansion around ${\bf K}$ by introducing ${\bf k} = {\bf K} + \delta{\bf k}$, with $| \delta {\bf k}| <\Lambda,~\Lambda \ll |{\bf K}|$, gives
\begin{eqnarray}
\nonumber H_{\bf K} \simeq \sum_{|\delta {\bf k}|<\Lambda} v_F |\delta{\bf k}| \bigg{[} && \psi^\dagger_{1R \downarrow}(\delta{\bf k}) \psi_{1R\downarrow}(\delta {\bf k}) - \\
&& \hspace{0.5cm} - \psi^\dagger_{2R\downarrow}(\delta {\bf k}) \psi_{2R\downarrow}(\delta {\bf k})\bigg{],}\label{Eq:H_K}
\end{eqnarray}
where we introduced $d^\downarrow_b({\bf K} + \delta {\bf k}) \equiv \psi_{bR\downarrow}(\delta {\bf k})$  with $R$ labeling the valley at ${\bf K}$ and $v_F \equiv \sqrt{3}t/2$ is the Fermi velocity of each band at ${\bf K}$. It is more convenient to transform the continuum fields defined above to real space, defining fields
\begin{eqnarray}
\psi_{bR\downarrow}({\bf r}) =\sqrt{\frac{1}{N_{uc}}} \sum_{|\delta{\bf k}|< \Lambda} e^{i \delta{\bf k}\cdot {\bf r}}\psi_{bR\downarrow}(\delta{\bf k}).
\end{eqnarray}
Therefore, in the low-energy description, we can re-express the spin-down fermion field as
\begin{eqnarray}
c_{\downarrow}({\bf r},a) \simeq \sum_{b=1,2} v_{b R\downarrow}(a) \psi_{b \downarrow}({\bf r})e^{i {\bf K}\cdot {\bf r}},
\end{eqnarray}
where we defined $v^{\downarrow}_b ({\bf K}+ \delta{\bf k},a) \equiv v_{bR\downarrow}(a)$.

Similarly, we can also obtain the low-energy description at ${\bf K'}$. At ${\bf K'}$, only the spin-up fermions are gapless and expansion around ${\bf K'}$ with small momentum shift $\delta {\bf k}$ gives
\begin{eqnarray}
\nonumber H_{\bf K'} \simeq \sum_{\delta {\bf k}<\Lambda} v_F |\delta {\bf k}|\bigg{[} &&\psi^\dagger_{1 L\uparrow}(\delta{\bf k}) \psi_{1L\uparrow}(\delta {\bf k}) - \\
&& \hspace{0.5cm} - \psi^\dagger_{2L\uparrow}(\delta {\bf k}) \psi_{2L\uparrow}(\delta {\bf k})\bigg{]}, \label{Eq:H_K'}
\end{eqnarray}
where similarly we defined $d^\uparrow_b({\bf K'} + \delta{\bf k}) \equiv \psi_{bL\uparrow}(\delta{\bf k})$. We can define a similar transformation to the real space as above, and therefore the spin-down fermion field can be effectively expressed as
\begin{eqnarray}
c_{\uparrow}({\bf r},a) \simeq \sum_{b=1,2} v_{b L\uparrow}(a)\psi_{bL\uparrow}({\bf r})e^{i{\bf K'}\cdot {\bf r}},
\end{eqnarray}
with $v^\uparrow_b({\bf K'} + \delta {\bf k},a) \equiv v_{b L\uparrow}(a)$, $L$ labeling the valley at ${\bf K'}$, and remember ${\bf K'} = - {\bf K}$. The action for the low-energy description is
\begin{eqnarray}
S_{0,P} = \int \frac{d^2{\bf q} d\omega}{(2\pi)^3} \left[\psi^\dagger_{b P\alpha_P} (q) (-i \omega) \psi_{b P\alpha_P} (q) + H_P\right],~~~~~
\end{eqnarray}
with $P = R/L = {\bf K}/{\bf K'}$ and $\alpha_{R/L} = \downarrow/\uparrow$ and we use $2+1$ dimensional vector $q$ representing frequency and momentum $(\omega,~{\bf q})$. We can also define the Green's functions as
\begin{eqnarray}
\nonumber \la \psi^\dagger_{bL\uparrow}(q) \psi_{bL\uparrow}( q') \ra &=& \la \psi^\dagger_{bR\downarrow}( q) \psi_{bR\downarrow}(q') \ra \\
&=& \frac{i \omega -(-1)^b v_F |{\bf q}|}{(i\omega)^2 - (v_F |{\bf q}|)^2} \delta^{(3)}_{q q'},
\end{eqnarray}
and we introduce the abbreviation $\delta^{(3)}_{q q'} = (2\pi)^3 \delta(\omega-\omega') \delta^{(2)}({\bf q} - {\bf q'})$.

In order to write down the general expression of the four-fermion interactions, we need first to obtain the symmetry transformation of the fields defined above. There are $S^z$-conservation, $U(1)$-charge, TRS, and $C_3$ in this system. Except TRS, the other else symmetry transformations are quite transparent. Let's focus on the symmetry transformation under TRS ($\mathcal{T}$), and we find
\begin{eqnarray}
&& v^{\uparrow *}_{bL}(a)  \mathcal{T} \psi^\uparrow_{bL} \mathcal{T}^{-1} = -v^\downarrow_{bR}(a)\psi^\downarrow_{bR},\label{Eq:TRS_1}\\
&& v^{\downarrow *}_{bR}(a) \mathcal{T} \psi^{\downarrow}_{bR} \mathcal{T}^{-1} =v^\uparrow_{bL}(a)\psi^\uparrow_{bL}.\label{Eq:TRS_2}
\end{eqnarray}
With TRS, the eigenvector-eigenvalue pairs have the property, $\vec{v}^\uparrow_b ({\bf k})=[\vec{v}^\downarrow_b (-{\bf k})]^*$  and $\epsilon^{\uparrow}_b ({\bf k}) = \epsilon^{\downarrow}_b (-{\bf k})$, which gives $v^{\uparrow *}_{b L}(a) = v^{\downarrow}_{bR}(a)$. We can use the properties above to simplify the TRS transformation in (\ref{Eq:TRS_1}) and (\ref{Eq:TRS_2}), but as far as the RG analysis presented below is concerned, we don't need to do that.

The general expressions of the local four-fermion interactions in terms of the continuum fields defined above are shown below. For simplicity in the expression, we define below $f_{b P \alpha}(a) \equiv v^\alpha_{b P} (a) \psi^\alpha_{b P}( {\bf r})$, and the local four-fermion action can be written as (repeated $a$ means summation over the eigenvector elements)
\begin{widetext}
\begin{eqnarray}
\nonumber S_{int} =~&& \omega^{a}_{11} f^\dagger_{1L\uparrow}(a) f_{1L\uparrow}(a) f^\dagger_{1R\downarrow}(a)f_{1R\downarrow}(a) + \omega^a_{22} f^\dagger_{2L\uparrow}(a) f_{2L\uparrow}(a) f^\dagger_{2R\downarrow}(a) f_{2R\downarrow}(a) +\\
\nonumber && + \omega^a_{12}\left[f^\dagger_{1L\uparrow}(a)f_{1L\uparrow}(a)f^\dagger_{2R\downarrow}(a)f_{2R\downarrow}(a) + f^\dagger_{1R\downarrow}(a)f_{1R\downarrow}(a)f^\dagger_{2L\uparrow}(a) f_{2L\uparrow}(a)\right]+\\
\nonumber && +\lambda^a_{12}\left[ (f^\dagger_{1L\uparrow}(a)f_{1L\uparrow}(a)f^\dagger_{1R\downarrow}(a)f_{2R\downarrow}(a) +f^\dagger_{1R\downarrow}(a)f_{1R\downarrow}(a)f^\dagger_{1L\uparrow}(a)f_{2L\uparrow}(a))+\Hc \right] + \\
\nonumber && + \lambda^a_{21}\left[(f^\dagger_{2L\uparrow}(a) f_{2L\uparrow}(a)f^\dagger_{1R\downarrow}(a)f_{2R\downarrow}(a)+f^\dagger_{2R\downarrow}f_{2R\downarrow}(a)f^\dagger_{1L\uparrow}(a)f_{2L\uparrow}(a))+\Hc \right] +\\
&& + u^a_{12}\left[ f^\dagger_{1L\uparrow}(a) f_{2L\uparrow}(a)f^\dagger_{1R\downarrow}(a)f_{2R\downarrow}(a) + \Hc \right]  + u^a_{21} \left[ f^\dagger_{1L\uparrow}(a)f_{2L\uparrow}(a) f^\dagger_{2R\downarrow}(a)f_{1R\downarrow}(a) + \Hc \right],
\end{eqnarray}
\end{widetext}
and we remark that in the presence of the on-site interaction $U$, Eq.~(\ref{Eq:Hubbard}), all the bare couplings above are simply equal to $U$.

The RG analysis in a nutshell is to integral out the fast-momentum modes defined within a momentum shell between $[\Lambda/b, \Lambda]$, with $b \equiv e^{d\ell}\simeq 1 + d\ell$ slightly bigger than one. The mathematical form at the tree-level is
\begin{eqnarray}
S_{eff,<} = \la S_{int} \ra_>,
\end{eqnarray}
where the subscript $>$ means momentum shell integral of the fast-momentum modes.

At the tree-level, we find the corrections are
\begin{widetext}
\begin{eqnarray}
\nonumber \la S_{int} \ra_>=  &&-\frac{\Lambda^2}{4\pi}d\ell  \bigg{[} \omega^a_{11} \big{|}v_{1R\downarrow}(a) \big{|}^2 - \omega^a_{12} \big{|} v_{2R\downarrow}(a) \big{|}^2 \bigg{]}f^\dagger_{1L\uparrow}(a)  f_{1 L \uparrow}(a)  + \frac{\Lambda^2}{4\pi}d\ell \bigg{[} \omega^a_{22} \big{|} v_{2R\downarrow}(a) \big{|}^2 - \omega^a_{12} \big{|} v_{1R\downarrow}(a) \big{|}^2 \bigg{]} f^\dagger_{2L \uparrow}(a) f_{2L\uparrow}(a) \\
\nonumber && - \frac{\Lambda^2}{4\pi}d\ell \bigg{[} \omega^a_{11} \big{|} v_{1L\uparrow}(a) \big{|}^2 - \omega^a_{12} \big{|} v_{2L \uparrow}(a) \big{|}^2 \bigg{]} f^\dagger_{1R \downarrow}(a) f_{1R\downarrow}(a) + \frac{\Lambda^2}{4\pi}d\ell \bigg{[} \omega^a_{22} \big{|} v_{2L \uparrow}(a) \big{|}^2 - \omega^a_{12} \big{|} v_{1L\uparrow}(a) \big{|}^2 \bigg{]} f^\dagger_{2R\downarrow}(a) f_{2R\downarrow}(a)\\
\nonumber && -\frac{\Lambda^2}{4\pi}d\ell  \bigg{[} \lambda^a_{12}\big{|} v_{1L\uparrow}(a) \big{|}^2  -  \lambda^a_{21} \big{|}v_{2L\uparrow}(a) \big{|}^2\bigg{]}\left( f^\dagger_{1R\downarrow}(a) f_{2R\downarrow}(a)+\Hc \right)\\
&& -\frac{\Lambda^2}{4\pi} d\ell \bigg{[}  \lambda^a_{12}\big{|}v_{1R\downarrow}(a) \big{|}^2 -  \lambda^a_{21}\big{|} v_{2R\downarrow}(a) \big{|}^2\bigg{]}\left(f^\dagger_{1L\uparrow}(a) f_{2L\uparrow}(a)+ \Hc \right), \label{Eq:tree-level_RG}
\end{eqnarray}
\end{widetext}
and all the four-fermion couplings are irrelevant,
\begin{eqnarray}
\frac{dg}{d\ell} = - g,
\end{eqnarray}
with $g= \omega^a$-s, $\lambda^a$-s, $u^a$-s introduced above and $\ell$ is the logarithm of the length scale in RG analysis..

The bare couplings of $\omega^a_{11}(\ell =0) = \omega^a_{22}(0) = \omega^a_{12} (0)= U$, and we numerically check that $ \big{|} v_{1R\downarrow}(a) \big{|}^2 = \big{|} v_{2R\downarrow}(a) \big{|}^2$, $\big{|} v_{1L\uparrow}(a)\big{|}^2 = \big{|} v_{2L\uparrow}(a)\big{|}^2$. At the tree-level RG analysis, all the couplings decays at the same rate under RG flow. Before all the four-fermion couplings flow to negligible values, at some small $\ell_c$, we have $\omega^a_{11} (\ell_c) = \omega^a_{22}(\ell_c) =\omega^a_{12}(\ell_c)=\lambda^a_{12}(\ell_c) = \lambda^a_{21}(\ell_c)$, and hence the bilinear corrections generated by these irrelevant four-fermion interactions completely cancel each other, which {\it leaves no corrections at the tree-level RG analysis.} Therefore, we conclude such long-wavelength analysis can not capture the shift of the boundary between the topologically trivial and nontrivial phases. The boundary shift can only be captured, at least in this model, by the {\it lattice} Hamiltonian which is not coarse-grained.

\section{Kane-Mele-U-V model in the presence of staggered potentials}\label{App:UV}
In this appendix, we will consider a more extended interaction which includes both the Hubbard $U$ and the nearest-neighbor $V$. The presence of the nearest-neighbor $V$ within Hartree-Fock contribute both the diagonal terms and the off-diagonal terms to the original Hamiltonian. The diagonal terms obviously renormalize the mass terms and the off-diagonal terms renormalize the nearest-neighbor hopping amplitude $t$, resulting in renormalizing the velocity of the Dirac fermions in the critical phase. Within the Hartree-Fock picture, besides the expectation values of the densities defined in the on-site Hubbard case, we also need to introduce
\begin{eqnarray}
&& \bigg{\la} c^\dagger_{\sigma}({\bf r}, B) c_\sigma({\bf r},A) \bigg{\ra} \equiv \left( \chi_\sigma \right)^*,\\
&& \bigg{\la} c^\dagger_{\sigma}({\bf r} + \vec{e}_a, B) c_\sigma({\bf r},A) \bigg{\ra} \equiv \left( \chi_{\sigma}(\vec{e}_a)\right)^*,
\end{eqnarray}
where $\vec{e}_{a = 1,2}$ defined in Fig.~\ref{Fig:honeycomb}. The nearest-neighbor $V$ contributes additional terms to the full Hamiltonian. In the matrix form, the additional terms can be expressed as
\begin{equation}
h_v({\bf k})=V \begin{pmatrix}
3\la n_B \ra & -\frak{f}_{\uparrow}({\bf k}) & 0 & 0\\
-(\frak{f}_{\uparrow}({\bf k}))^* & 3 \la n_A \ra & 0 & 0\\
0 & 0 & 3 \la n_B \ra & - \frak{f}_{\downarrow}({\bf k})\\
0 & 0 & - ( \frak{f}_{ \downarrow}({\bf k}) )^* & 3 \la n_A \ra
\end{pmatrix},
\end{equation}
where $\frak{f}_{\sigma}({\bf k}) \equiv (\chi_{\sigma} )^* + e^{i {\bf k}\cdot \vec{e}_1} (\chi_\sigma (\vec{e}_1))^* + e^{i {\bf k}\cdot \vec{e}_2} (\chi_\sigma (\vec{e}_2))^*$. By $C_3$ symmetry, we can simplify the result by identifying $\chi_\sigma = \chi_\sigma(\vec{e}_1) = \chi_\sigma(\vec{e}_2)$. We can see that $f_{\sigma}({\bf k})$ is proportional to $f({\bf k})$ defined in Eq.~\ref{Eq:h_KMs} and therefore vanish at momentums ${\bf K}$ and ${\bf K'}$. We also set $\lambda^A_m = \lambda^B_m = \lambda_m$ for simplicity.
\begin{figure}[t]
\subfigure[]
{\label{Fig:UV_repulsive} \includegraphics[width=1.6in]{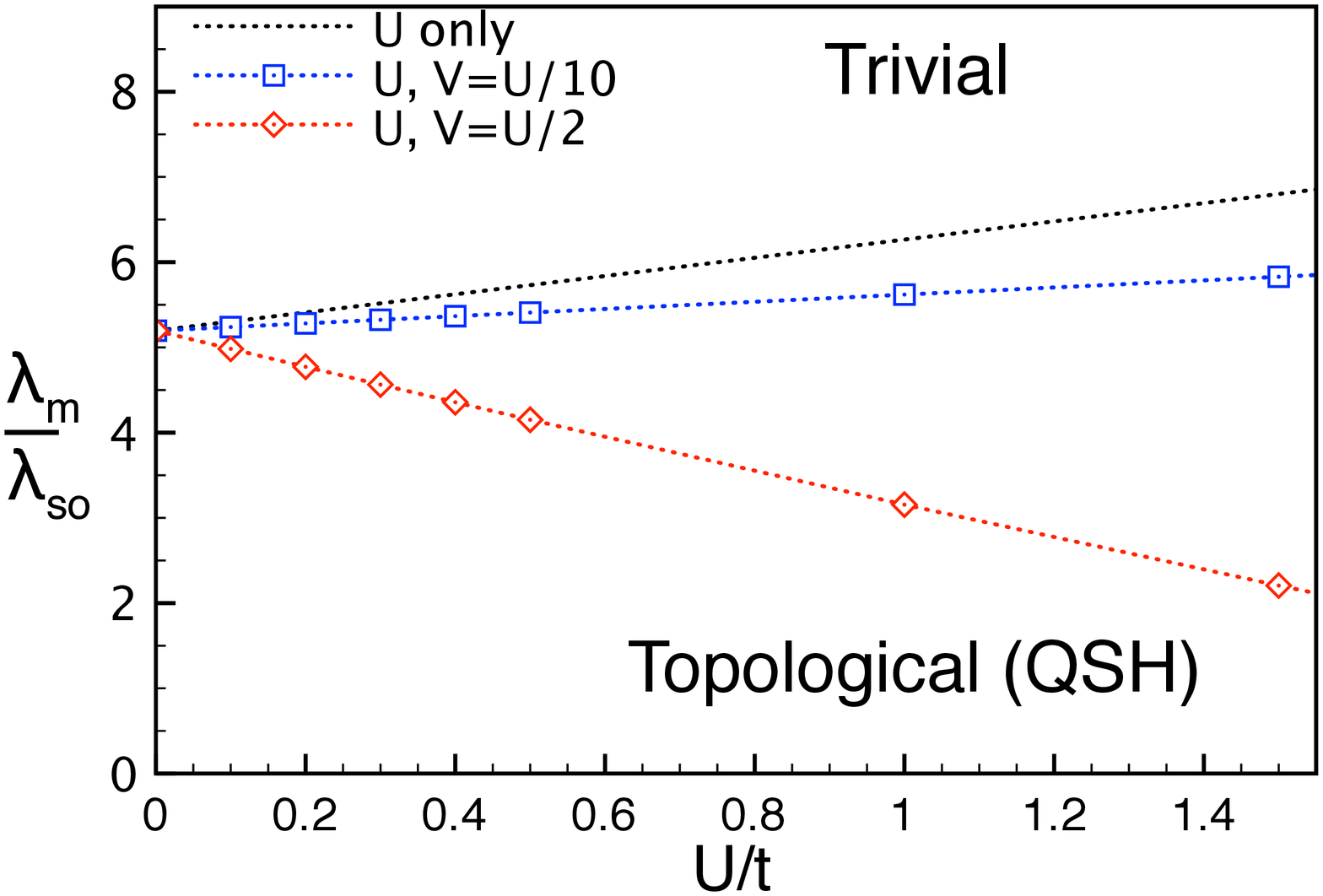}}
\subfigure[]
{\label{Fig:UV_attractive} \includegraphics[width=1.6in]{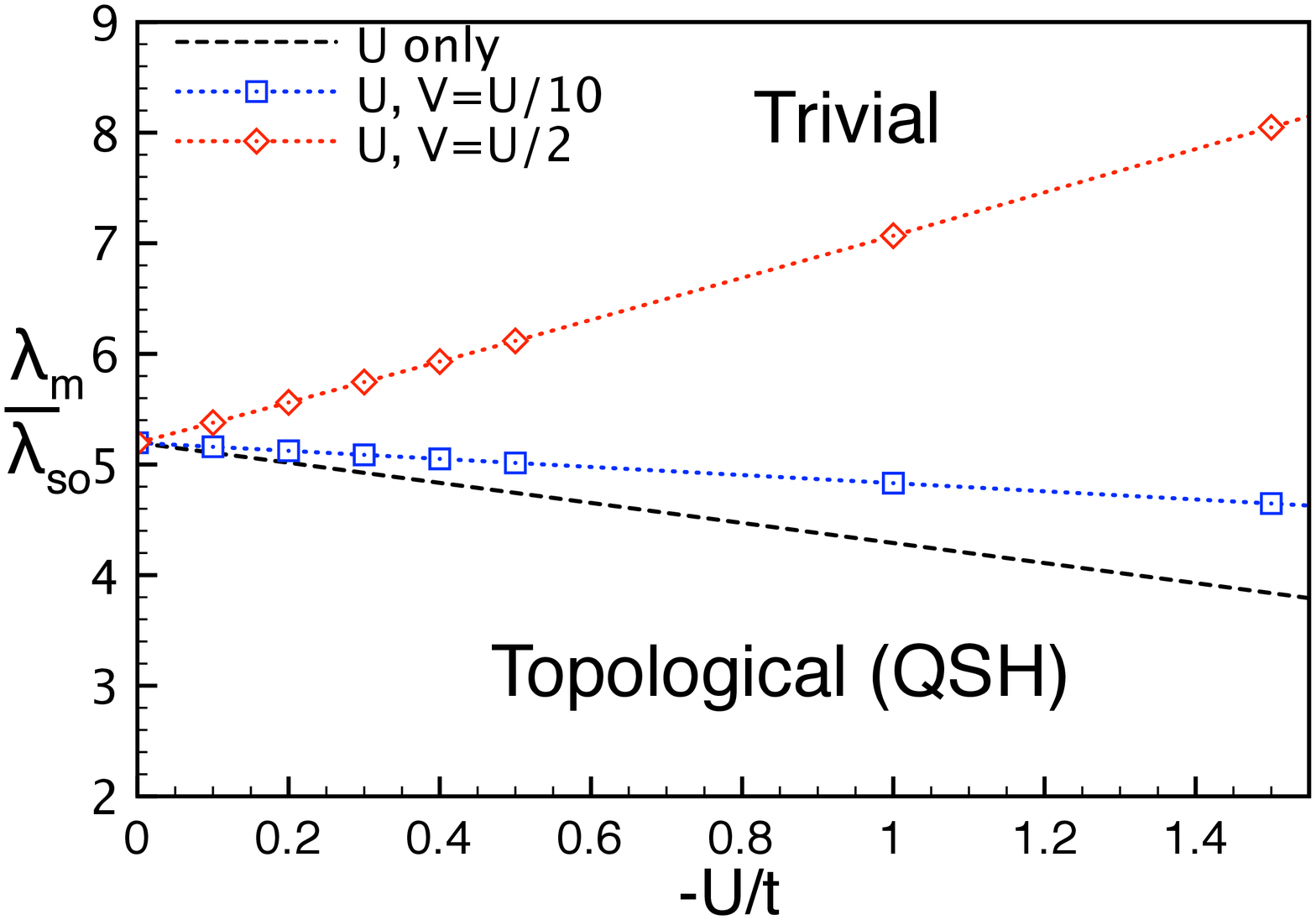}}
\caption{Illustration of the phase boundary between trivial and nontrivial phase in the present of both U and V repulsion. For illustration, we take $t=1$, $\lambda_{so} =0.4$, and choose $V=U/2$ and $V=U/10$. The blue open dot squares represent the boundary in the case of $V=U/10$ and the open red diamonds line represents the boundary in the case of $V=U/10$. The black line represents the boundary in the present of on-site Hubbard $U$. (a) The repulsive interaction case, $U, V >0$. We can see that in this case. If the nearest-neighbor $V$ is much smaller than $U$, $V=U/10$, the interactions still tend to stabilize the topological phase. But due to the competition between $U$ and $V$ the topological widow is less widened by the interactions. If $V$ is more comparable to $U$,$V=U/2$ in this case, the effects of $V$ will be dominant over $U$ and will tend to destabilize the topological phase. According to the gap function defined around momentum ${\bf K}$ point, Eq.~(\ref{Eq:UV_gap}), the transition point is at $U= 6V$. (b) The attractive interaction case, $U, V <0$. The results in this case are qualitatively opposite to those in the repulsive case. When $V$ is more comparable to the $U$, the interactions tend to stabilize the topological phase.}
\label{Fig:UV_data}
\end{figure}
Focusing on momentum ${\bf K}$, we can see that the two of the four bands with the eigenvalues $E_{1/2}=\pm \lambda_m \mp 2\lambda_{so}g(K) + \frac{U}{2} \la n_{A/B} \ra + 3V\la n_{B/A} \ra$ can be inverted due to the tuning of the ratio of $\lambda_m$ and $\lambda_{so}$. Therefore, we define the gap function as
\begin{eqnarray}
\nonumber && \Delta({\bf K}) = E_1 - E_2 \\
 && =  2 \lambda_m - 4 \lambda_{so} g(K) + \left( \frac{U}{2}-3V\right) \big{[} \la n_A \ra - \la n_B \ra \big{]}.\label{Eq:UV_gap}
\end{eqnarray}
Due to the presence of the staggered potentials, the density at B is larger than that at A, $\la n_B \ra > \la n_A \ra$, and the sign of the correction of the last term in the gap function depends on the competition between $U$ and $V$. For repulsive $U, V >0$, if $U>6V$, the last term is negative and the repulsive interactions stabilizes the topological phase. However, if $U<6V$, the last term is positive and then the interactions destabilizes the topological phase. On the other hand, the attractive $U, V <0$ would give the opposite results to the repulsive case.

For illustration, we numerically check the cases with $t =1$, $\lambda_{so} =0.4$, and choose $V=U/10$ and $V=U/2$ on a honeycomb lattice consisting of $200\times 200$ unit cells. The results on shown in Fig.~\ref{Fig:UV_data}. The blue open dot squares represent the boundary in the case of $V=U/10$ and the open red diamonds line represents the boundary in the case of $V=U/10$. The black dashed line represents the boundary in the present of only on-site Hubbard $U$ shown in Fig.~\ref{Fig:onsite_U}. Qualitatively, the results between the repulsive and the attractive case are opposite. In the repulsive interaction case, the more short-ranged the repulsive interactions are, the more stable the topological is. On the other hand, in the attractive interaction case, the more extended the attractive interactions are, the more stable the topological phase is.


\bibliography{biblio4KM}

\begin{thebibliography}{39}%
\makeatletter
\providecommand \@ifxundefined [1]{%
 \@ifx{#1\undefined}
}%
\providecommand \@ifnum [1]{%
 \ifnum #1\expandafter \@firstoftwo
 \else \expandafter \@secondoftwo
 \fi
}%
\providecommand \@ifx [1]{%
 \ifx #1\expandafter \@firstoftwo
 \else \expandafter \@secondoftwo
 \fi
}%
\providecommand \natexlab [1]{#1}%
\providecommand \enquote  [1]{``#1''}%
\providecommand \bibnamefont  [1]{#1}%
\providecommand \bibfnamefont [1]{#1}%
\providecommand \citenamefont [1]{#1}%
\providecommand \href@noop [0]{\@secondoftwo}%
\providecommand \href [0]{\begingroup \@sanitize@url \@href}%
\providecommand \@href[1]{\@@startlink{#1}\@@href}%
\providecommand \@@href[1]{\endgroup#1\@@endlink}%
\providecommand \@sanitize@url [0]{\catcode `\\12\catcode `\$12\catcode
  `\&12\catcode `\#12\catcode `\^12\catcode `\_12\catcode `\%12\relax}%
\providecommand \@@startlink[1]{}%
\providecommand \@@endlink[0]{}%
\providecommand \url  [0]{\begingroup\@sanitize@url \@url }%
\providecommand \@url [1]{\endgroup\@href {#1}{\urlprefix }}%
\providecommand \urlprefix  [0]{URL }%
\providecommand \Eprint [0]{\href }%
\providecommand \doibase [0]{http://dx.doi.org/}%
\providecommand \selectlanguage [0]{\@gobble}%
\providecommand \bibinfo  [0]{\@secondoftwo}%
\providecommand \bibfield  [0]{\@secondoftwo}%
\providecommand \translation [1]{[#1]}%
\providecommand \BibitemOpen [0]{}%
\providecommand \bibitemStop [0]{}%
\providecommand \bibitemNoStop [0]{.\EOS\space}%
\providecommand \EOS [0]{\spacefactor3000\relax}%
\providecommand \BibitemShut  [1]{\csname bibitem#1\endcsname}%
\let\auto@bib@innerbib\@empty
\bibitem [{\citenamefont {Fu}\ and\ \citenamefont {Kane}(2007)}]{fu2007}%
  \BibitemOpen
  \bibfield  {author} {\bibinfo {author} {\bibfnamefont {L.}~\bibnamefont
  {Fu}}\ and\ \bibinfo {author} {\bibfnamefont {C.~L.}\ \bibnamefont {Kane}},\
  }\href {\doibase 10.1103/PhysRevB.76.045302} {\bibfield  {journal} {\bibinfo
  {journal} {Phys. Rev. B}\ }\textbf {\bibinfo {volume} {76}},\ \bibinfo
  {pages} {045302} (\bibinfo {year} {2007})}\BibitemShut {NoStop}%
\bibitem [{\citenamefont {Moore}\ and\ \citenamefont
  {Balents}(2007)}]{moore2007}%
  \BibitemOpen
  \bibfield  {author} {\bibinfo {author} {\bibfnamefont {J.~E.}\ \bibnamefont
  {Moore}}\ and\ \bibinfo {author} {\bibfnamefont {L.}~\bibnamefont
  {Balents}},\ }\href@noop {} {\bibfield  {journal} {\bibinfo  {journal} {Phys.
  Rev. B}\ }\textbf {\bibinfo {volume} {75}},\ \bibinfo {pages} {121306}
  (\bibinfo {year} {2007})}\BibitemShut {NoStop}%
\bibitem [{\citenamefont {Roy}(2009)}]{roy2009}%
  \BibitemOpen
  \bibfield  {author} {\bibinfo {author} {\bibfnamefont {R.}~\bibnamefont
  {Roy}},\ }\href@noop {} {\bibfield  {journal} {\bibinfo  {journal} {Phys.
  Rev. B}\ }\textbf {\bibinfo {volume} {79}},\ \bibinfo {pages} {195322}
  (\bibinfo {year} {2009})}\BibitemShut {NoStop}%
\bibitem [{\citenamefont {Hasan}\ and\ \citenamefont {Kane}(2010)}]{hasan2010}%
  \BibitemOpen
  \bibfield  {author} {\bibinfo {author} {\bibfnamefont {M.~Z.}\ \bibnamefont
  {Hasan}}\ and\ \bibinfo {author} {\bibfnamefont {C.~L.}\ \bibnamefont
  {Kane}},\ }\href@noop {} {\bibfield  {journal} {\bibinfo  {journal} {Rev.
  Mod. Phys.}\ }\textbf {\bibinfo {volume} {82}},\ \bibinfo {pages} {3045}
  (\bibinfo {year} {2010})}\BibitemShut {NoStop}%
\bibitem [{\citenamefont {Qi}\ and\ \citenamefont {Zhang}(2011)}]{qi2011}%
  \BibitemOpen
  \bibfield  {author} {\bibinfo {author} {\bibfnamefont {X.-L.}\ \bibnamefont
  {Qi}}\ and\ \bibinfo {author} {\bibfnamefont {S.-C.}\ \bibnamefont {Zhang}},\
  }\href@noop {} {\bibfield  {journal} {\bibinfo  {journal} {Rev. Mod. Phys.}\
  }\textbf {\bibinfo {volume} {83}},\ \bibinfo {pages} {1057} (\bibinfo {year}
  {2011})}\BibitemShut {NoStop}%
\bibitem [{\citenamefont {Pankratov}\ \emph {et~al.}(1987)\citenamefont
  {Pankratov}, \citenamefont {Pakhomov},\ and\ \citenamefont
  {Volkov}}]{Pankratov1987}%
  \BibitemOpen
  \bibfield  {author} {\bibinfo {author} {\bibfnamefont {O.}~\bibnamefont
  {Pankratov}}, \bibinfo {author} {\bibfnamefont {S.}~\bibnamefont {Pakhomov}},
  \ and\ \bibinfo {author} {\bibfnamefont {B.}~\bibnamefont {Volkov}},\ }\href
  {\doibase http://dx.doi.org/10.1016/0038-1098(87)90934-3} {\bibfield
  {journal} {\bibinfo  {journal} {Solid State Communications}\ }\textbf
  {\bibinfo {volume} {61}},\ \bibinfo {pages} {93 } (\bibinfo {year}
  {1987})}\BibitemShut {NoStop}%
\bibitem [{\citenamefont {Bernevig}\ \emph {et~al.}(2006)\citenamefont
  {Bernevig}, \citenamefont {Hughes},\ and\ \citenamefont
  {Zhang}}]{Bernevig2006}%
  \BibitemOpen
  \bibfield  {author} {\bibinfo {author} {\bibfnamefont {B.~A.}\ \bibnamefont
  {Bernevig}}, \bibinfo {author} {\bibfnamefont {T.~L.}\ \bibnamefont
  {Hughes}}, \ and\ \bibinfo {author} {\bibfnamefont {S.-C.}\ \bibnamefont
  {Zhang}},\ }\href@noop {} {\bibfield  {journal} {\bibinfo  {journal}
  {Science}\ }\textbf {\bibinfo {volume} {314}},\ \bibinfo {pages} {1757}
  (\bibinfo {year} {2006})}\BibitemShut {NoStop}%
\bibitem [{\citenamefont {Kšnig}\ \emph {et~al.}(2007)\citenamefont {Kšnig},
  \citenamefont {Wiedmann}, \citenamefont {BrŸne}, \citenamefont {Roth},
  \citenamefont {Buhmann}, \citenamefont {Molenkamp}, \citenamefont {Qi},\ and\
  \citenamefont {Zhang}}]{Kšnig2007}%
  \BibitemOpen
  \bibfield  {author} {\bibinfo {author} {\bibfnamefont {M.}~\bibnamefont
  {Kšnig}}, \bibinfo {author} {\bibfnamefont {S.}~\bibnamefont {Wiedmann}},
  \bibinfo {author} {\bibfnamefont {C.}~\bibnamefont {BrŸne}}, \bibinfo
  {author} {\bibfnamefont {A.}~\bibnamefont {Roth}}, \bibinfo {author}
  {\bibfnamefont {H.}~\bibnamefont {Buhmann}}, \bibinfo {author} {\bibfnamefont
  {L.~W.}\ \bibnamefont {Molenkamp}}, \bibinfo {author} {\bibfnamefont {X.-L.}\
  \bibnamefont {Qi}}, \ and\ \bibinfo {author} {\bibfnamefont {S.-C.}\
  \bibnamefont {Zhang}},\ }\href {\doibase 10.1126/science.1148047} {\bibfield
  {journal} {\bibinfo  {journal} {Science}\ }\textbf {\bibinfo {volume}
  {318}},\ \bibinfo {pages} {766} (\bibinfo {year} {2007})}\BibitemShut
  {NoStop}%
\bibitem [{\citenamefont {Kane}\ and\ \citenamefont
  {Mele}(2005{\natexlab{a}})}]{kane2005a}%
  \BibitemOpen
  \bibfield  {author} {\bibinfo {author} {\bibfnamefont {C.~L.}\ \bibnamefont
  {Kane}}\ and\ \bibinfo {author} {\bibfnamefont {E.~J.}\ \bibnamefont
  {Mele}},\ }\href {\doibase 10.1103/PhysRevLett.95.226801} {\bibfield
  {journal} {\bibinfo  {journal} {Phys. Rev. Lett.}\ }\textbf {\bibinfo
  {volume} {95}},\ \bibinfo {pages} {226801} (\bibinfo {year}
  {2005}{\natexlab{a}})}\BibitemShut {NoStop}%
\bibitem [{\citenamefont {Haldane}(1988)}]{haldane1988}%
  \BibitemOpen
  \bibfield  {author} {\bibinfo {author} {\bibfnamefont {F.~D.~M.}\
  \bibnamefont {Haldane}},\ }\href {\doibase 10.1103/PhysRevLett.61.2015}
  {\bibfield  {journal} {\bibinfo  {journal} {Phys. Rev. Lett.}\ }\textbf
  {\bibinfo {volume} {61}},\ \bibinfo {pages} {2015} (\bibinfo {year}
  {1988})}\BibitemShut {NoStop}%
\bibitem [{\citenamefont {Rachel}\ and\ \citenamefont {{Le
  Hur}}(2010)}]{rachel2010}%
  \BibitemOpen
  \bibfield  {author} {\bibinfo {author} {\bibfnamefont {S.}~\bibnamefont
  {Rachel}}\ and\ \bibinfo {author} {\bibfnamefont {K.}~\bibnamefont {{Le
  Hur}}},\ }\href {\doibase 10.1103/PhysRevB.82.075106} {\bibfield  {journal}
  {\bibinfo  {journal} {Phys. Rev. B}\ }\textbf {\bibinfo {volume} {82}},\
  \bibinfo {pages} {075106} (\bibinfo {year} {2010})}\BibitemShut {NoStop}%
\bibitem [{\citenamefont {Yu}\ \emph {et~al.}(2011)\citenamefont {Yu},
  \citenamefont {Xie},\ and\ \citenamefont {Li}}]{yu2011}%
  \BibitemOpen
  \bibfield  {author} {\bibinfo {author} {\bibfnamefont {S.-L.}\ \bibnamefont
  {Yu}}, \bibinfo {author} {\bibfnamefont {X.~C.}\ \bibnamefont {Xie}}, \ and\
  \bibinfo {author} {\bibfnamefont {J.-X.}\ \bibnamefont {Li}},\ }\href@noop {}
  {\bibfield  {journal} {\bibinfo  {journal} {Phys. Rev. Lett.}\ }\textbf
  {\bibinfo {volume} {107}},\ \bibinfo {pages} {010401} (\bibinfo {year}
  {2011})}\BibitemShut {NoStop}%
\bibitem [{\citenamefont {Zheng}\ \emph {et~al.}(2011)\citenamefont {Zheng},
  \citenamefont {Zhang},\ and\ \citenamefont {Wu}}]{zheng2011}%
  \BibitemOpen
  \bibfield  {author} {\bibinfo {author} {\bibfnamefont {D.}~\bibnamefont
  {Zheng}}, \bibinfo {author} {\bibfnamefont {G.-M.}\ \bibnamefont {Zhang}}, \
  and\ \bibinfo {author} {\bibfnamefont {C.}~\bibnamefont {Wu}},\ }\href@noop
  {} {\bibfield  {journal} {\bibinfo  {journal} {Phys. Rev. B}\ }\textbf
  {\bibinfo {volume} {84}},\ \bibinfo {pages} {205121} (\bibinfo {year}
  {2011})}\BibitemShut {NoStop}%
\bibitem [{\citenamefont {Hohenadler}\ \emph {et~al.}(2011)\citenamefont
  {Hohenadler}, \citenamefont {Lang},\ and\ \citenamefont
  {Assaad}}]{hohenadler2011}%
  \BibitemOpen
  \bibfield  {author} {\bibinfo {author} {\bibfnamefont {M.}~\bibnamefont
  {Hohenadler}}, \bibinfo {author} {\bibfnamefont {T.~C.}\ \bibnamefont
  {Lang}}, \ and\ \bibinfo {author} {\bibfnamefont {F.~F.}\ \bibnamefont
  {Assaad}},\ }\href@noop {} {\bibfield  {journal} {\bibinfo  {journal} {Phys.
  Rev. Lett.}\ }\textbf {\bibinfo {volume} {106}},\ \bibinfo {pages} {100403}
  (\bibinfo {year} {2011})}\BibitemShut {NoStop}%
\bibitem [{\citenamefont {Budich}\ \emph {et~al.}(2012)\citenamefont {Budich},
  \citenamefont {Thomale}, \citenamefont {Li}, \citenamefont {Laubach},\ and\
  \citenamefont {Zhang}}]{Budich:prb12}%
  \BibitemOpen
  \bibfield  {author} {\bibinfo {author} {\bibfnamefont {J.~C.}\ \bibnamefont
  {Budich}}, \bibinfo {author} {\bibfnamefont {R.}~\bibnamefont {Thomale}},
  \bibinfo {author} {\bibfnamefont {G.}~\bibnamefont {Li}}, \bibinfo {author}
  {\bibfnamefont {M.}~\bibnamefont {Laubach}}, \ and\ \bibinfo {author}
  {\bibfnamefont {S.-C.}\ \bibnamefont {Zhang}},\ }\href {\doibase
  10.1103/PhysRevB.86.201407} {\bibfield  {journal} {\bibinfo  {journal} {Phys.
  Rev. B}\ }\textbf {\bibinfo {volume} {86}},\ \bibinfo {pages} {201407}
  (\bibinfo {year} {2012})}\BibitemShut {NoStop}%
\bibitem [{\citenamefont {Wu}\ \emph {et~al.}(2012)\citenamefont {Wu},
  \citenamefont {Rachel}, \citenamefont {Liu},\ and\ \citenamefont {{Le
  Hur}}}]{wuwei2012}%
  \BibitemOpen
  \bibfield  {author} {\bibinfo {author} {\bibfnamefont {W.}~\bibnamefont
  {Wu}}, \bibinfo {author} {\bibfnamefont {S.}~\bibnamefont {Rachel}}, \bibinfo
  {author} {\bibfnamefont {W.-M.}\ \bibnamefont {Liu}}, \ and\ \bibinfo
  {author} {\bibfnamefont {K.}~\bibnamefont {{Le Hur}}},\ }\href@noop {}
  {\bibfield  {journal} {\bibinfo  {journal} {Phys. Rev. B}\ }\textbf {\bibinfo
  {volume} {85}},\ \bibinfo {pages} {205102} (\bibinfo {year}
  {2012})}\BibitemShut {NoStop}%
\bibitem [{\citenamefont {Hohenadler}\ \emph {et~al.}(2012)\citenamefont
  {Hohenadler}, \citenamefont {Meng}, \citenamefont {Lang}, \citenamefont
  {Wessel}, \citenamefont {Muramatsu},\ and\ \citenamefont
  {Assaad}}]{hohenadler2012}%
  \BibitemOpen
  \bibfield  {author} {\bibinfo {author} {\bibfnamefont {M.}~\bibnamefont
  {Hohenadler}}, \bibinfo {author} {\bibfnamefont {Z.~Y.}\ \bibnamefont
  {Meng}}, \bibinfo {author} {\bibfnamefont {T.~C.}\ \bibnamefont {Lang}},
  \bibinfo {author} {\bibfnamefont {S.}~\bibnamefont {Wessel}}, \bibinfo
  {author} {\bibfnamefont {A.}~\bibnamefont {Muramatsu}}, \ and\ \bibinfo
  {author} {\bibfnamefont {F.~F.}\ \bibnamefont {Assaad}},\ }\href@noop {}
  {\bibfield  {journal} {\bibinfo  {journal} {Phys. Rev. B}\ }\textbf {\bibinfo
  {volume} {85}},\ \bibinfo {pages} {115132} (\bibinfo {year}
  {2012})}\BibitemShut {NoStop}%
\bibitem [{\citenamefont {Griset}\ and\ \citenamefont {Xu}(2012)}]{Griset2012}%
  \BibitemOpen
  \bibfield  {author} {\bibinfo {author} {\bibfnamefont {C.}~\bibnamefont
  {Griset}}\ and\ \bibinfo {author} {\bibfnamefont {C.}~\bibnamefont {Xu}},\
  }\href {\doibase 10.1103/PhysRevB.85.045123} {\bibfield  {journal} {\bibinfo
  {journal} {Phys. Rev. B}\ }\textbf {\bibinfo {volume} {85}},\ \bibinfo
  {pages} {045123} (\bibinfo {year} {2012})}\BibitemShut {NoStop}%
\bibitem [{\citenamefont {Hung}\ \emph {et~al.}(2013)\citenamefont {Hung},
  \citenamefont {Wang}, \citenamefont {Gu},\ and\ \citenamefont
  {Fiete}}]{Hung2013}%
  \BibitemOpen
  \bibfield  {author} {\bibinfo {author} {\bibfnamefont {H.-H.}\ \bibnamefont
  {Hung}}, \bibinfo {author} {\bibfnamefont {L.}~\bibnamefont {Wang}}, \bibinfo
  {author} {\bibfnamefont {Z.-C.}\ \bibnamefont {Gu}}, \ and\ \bibinfo {author}
  {\bibfnamefont {G.~A.}\ \bibnamefont {Fiete}},\ }\href@noop {} {\bibfield
  {journal} {\bibinfo  {journal} {Phys. Rev. B}\ }\textbf {\bibinfo {volume}
  {87}},\ \bibinfo {pages} {121113} (\bibinfo {year} {2013})}\BibitemShut
  {NoStop}%
\bibitem [{\citenamefont {Hung}\ \emph {et~al.}(shed)\citenamefont {Hung},
  \citenamefont {Chua}, \citenamefont {Wang},\ and\ \citenamefont
  {Fiete}}]{hung2013b}%
  \BibitemOpen
  \bibfield  {author} {\bibinfo {author} {\bibfnamefont {H.-H.}\ \bibnamefont
  {Hung}}, \bibinfo {author} {\bibfnamefont {V.}~\bibnamefont {Chua}}, \bibinfo
  {author} {\bibfnamefont {L.}~\bibnamefont {Wang}}, \ and\ \bibinfo {author}
  {\bibfnamefont {G.~A.}\ \bibnamefont {Fiete}},\ }\href@noop {} {\bibfield
  {journal} {\bibinfo  {journal} {arXiv:1307.2659}\ } (\bibinfo {year}
  {unpublished})}\BibitemShut {NoStop}%
\bibitem [{\citenamefont {Lang}\ \emph {et~al.}(2013)\citenamefont {Lang},
  \citenamefont {Essin}, \citenamefont {Gurarie},\ and\ \citenamefont
  {Wessel}}]{lang2013}%
  \BibitemOpen
  \bibfield  {author} {\bibinfo {author} {\bibfnamefont {T.~C.}\ \bibnamefont
  {Lang}}, \bibinfo {author} {\bibfnamefont {A.~M.}\ \bibnamefont {Essin}},
  \bibinfo {author} {\bibfnamefont {V.}~\bibnamefont {Gurarie}}, \ and\
  \bibinfo {author} {\bibfnamefont {S.}~\bibnamefont {Wessel}},\ }\href@noop {}
  {\bibfield  {journal} {\bibinfo  {journal} {Phys. Rev. B}\ }\textbf {\bibinfo
  {volume} {87}},\ \bibinfo {pages} {205101} (\bibinfo {year}
  {2013})}\BibitemShut {NoStop}%
\bibitem [{\citenamefont {Zong}\ \emph {et~al.}(2013)\citenamefont {Zong},
  \citenamefont {He},\ and\ \citenamefont {Kou}}]{Zong2013}%
  \BibitemOpen
  \bibfield  {author} {\bibinfo {author} {\bibfnamefont {Y.-H.}\ \bibnamefont
  {Zong}}, \bibinfo {author} {\bibfnamefont {J.}~\bibnamefont {He}}, \ and\
  \bibinfo {author} {\bibfnamefont {S.-P.}\ \bibnamefont {Kou}},\ }\href
  {\doibase 10.1140/epjb/e2012-30514-3} {\bibfield  {journal} {\bibinfo
  {journal} {The European Physical Journal B}\ }\textbf {\bibinfo {volume}
  {86}},\ \bibinfo {pages} {1} (\bibinfo {year} {2013})}\BibitemShut {NoStop}%
\bibitem [{\citenamefont {Cai}\ \emph {et~al.}(2008)\citenamefont {Cai},
  \citenamefont {Chen}, \citenamefont {Kou},\ and\ \citenamefont
  {Wang}}]{Cai2008}%
  \BibitemOpen
  \bibfield  {author} {\bibinfo {author} {\bibfnamefont {Z.}~\bibnamefont
  {Cai}}, \bibinfo {author} {\bibfnamefont {S.}~\bibnamefont {Chen}}, \bibinfo
  {author} {\bibfnamefont {S.}~\bibnamefont {Kou}}, \ and\ \bibinfo {author}
  {\bibfnamefont {Y.}~\bibnamefont {Wang}},\ }\href@noop {} {\bibfield
  {journal} {\bibinfo  {journal} {Phys. Rev. B}\ }\textbf {\bibinfo {volume}
  {78}},\ \bibinfo {pages} {035123} (\bibinfo {year} {2008})}\BibitemShut
  {NoStop}%
\bibitem [{\citenamefont {Yuan}\ \emph {et~al.}(2012)\citenamefont {Yuan},
  \citenamefont {Gao}, \citenamefont {Chen}, \citenamefont {Ye}, \citenamefont
  {Zhou},\ and\ \citenamefont {Zhang}}]{Yuan2012}%
  \BibitemOpen
  \bibfield  {author} {\bibinfo {author} {\bibfnamefont {J.}~\bibnamefont
  {Yuan}}, \bibinfo {author} {\bibfnamefont {J.-H.}\ \bibnamefont {Gao}},
  \bibinfo {author} {\bibfnamefont {W.-Q.}\ \bibnamefont {Chen}}, \bibinfo
  {author} {\bibfnamefont {F.}~\bibnamefont {Ye}}, \bibinfo {author}
  {\bibfnamefont {Y.}~\bibnamefont {Zhou}}, \ and\ \bibinfo {author}
  {\bibfnamefont {F.-C.}\ \bibnamefont {Zhang}},\ }\href {\doibase
  10.1103/PhysRevB.86.104505} {\bibfield  {journal} {\bibinfo  {journal} {Phys.
  Rev. B}\ }\textbf {\bibinfo {volume} {86}},\ \bibinfo {pages} {104505}
  (\bibinfo {year} {2012})}\BibitemShut {NoStop}%
\bibitem [{\citenamefont {Hirsch}(1985)}]{hirsch1985}%
  \BibitemOpen
  \bibfield  {author} {\bibinfo {author} {\bibfnamefont {J.~E.}\ \bibnamefont
  {Hirsch}},\ }\href@noop {} {\bibfield  {journal} {\bibinfo  {journal} {Phys.
  Rev. B}\ }\textbf {\bibinfo {volume} {31}},\ \bibinfo {pages} {4403}
  (\bibinfo {year} {1985})}\BibitemShut {NoStop}%
\bibitem [{\citenamefont {Sorella}\ \emph {et~al.}(1989)\citenamefont
  {Sorella}, \citenamefont {Baroni}, \citenamefont {Car},\ and\ \citenamefont
  {Parrinello}}]{sorella1989}%
  \BibitemOpen
  \bibfield  {author} {\bibinfo {author} {\bibfnamefont {S.}~\bibnamefont
  {Sorella}}, \bibinfo {author} {\bibfnamefont {S.}~\bibnamefont {Baroni}},
  \bibinfo {author} {\bibfnamefont {R.}~\bibnamefont {Car}}, \ and\ \bibinfo
  {author} {\bibfnamefont {M.}~\bibnamefont {Parrinello}},\ }\href@noop {}
  {\bibfield  {journal} {\bibinfo  {journal} {Europhys. Lett.}\ }\textbf
  {\bibinfo {volume} {8}},\ \bibinfo {pages} {663} (\bibinfo {year}
  {1989})}\BibitemShut {NoStop}%
\bibitem [{\citenamefont {Assaad}(2002)}]{assaad2002}%
  \BibitemOpen
  \bibfield  {author} {\bibinfo {author} {\bibfnamefont {F.~F.}\ \bibnamefont
  {Assaad}},\ }\href@noop {} {\emph {\bibinfo {title} {Quantum {Monte Carlo}
  methods on lattices: The determinantal approach in Quantum Simulations of
  Complex Many-Body Systems: From Theory to Algorithms, Lecture Notes}}}\
  (\bibinfo  {publisher} {NIC Series Vol. {\bf 10}},\ \bibinfo {year}
  {2002})\BibitemShut {NoStop}%
\bibitem [{\citenamefont {Wang}\ and\ \citenamefont
  {Zhang}(2012{\natexlab{a}})}]{WangPRB}%
  \BibitemOpen
  \bibfield  {author} {\bibinfo {author} {\bibfnamefont {Z.}~\bibnamefont
  {Wang}}\ and\ \bibinfo {author} {\bibfnamefont {S.-C.}\ \bibnamefont
  {Zhang}},\ }\href {\doibase 10.1103/PhysRevB.86.165116} {\bibfield  {journal}
  {\bibinfo  {journal} {Phys. Rev. B}\ }\textbf {\bibinfo {volume} {86}},\
  \bibinfo {pages} {165116} (\bibinfo {year} {2012}{\natexlab{a}})}\BibitemShut
  {NoStop}%
\bibitem [{\citenamefont {Wang}\ and\ \citenamefont
  {Zhang}(2012{\natexlab{b}})}]{Wang_PRX}%
  \BibitemOpen
  \bibfield  {author} {\bibinfo {author} {\bibfnamefont {Z.}~\bibnamefont
  {Wang}}\ and\ \bibinfo {author} {\bibfnamefont {S.-C.}\ \bibnamefont
  {Zhang}},\ }\href {\doibase 10.1103/PhysRevX.2.031008} {\bibfield  {journal}
  {\bibinfo  {journal} {Phys. Rev. X}\ }\textbf {\bibinfo {volume} {2}},\
  \bibinfo {pages} {031008} (\bibinfo {year} {2012}{\natexlab{b}})}\BibitemShut
  {NoStop}%
\bibitem [{\citenamefont {Wang}\ and\ \citenamefont {Yan}(2013)}]{Wang2013}%
  \BibitemOpen
  \bibfield  {author} {\bibinfo {author} {\bibfnamefont {Z.}~\bibnamefont
  {Wang}}\ and\ \bibinfo {author} {\bibfnamefont {B.}~\bibnamefont {Yan}},\
  }\href {http://stacks.iop.org/0953-8984/25/i=15/a=155601} {\bibfield
  {journal} {\bibinfo  {journal} {Journal of Physics: Condensed Matter}\
  }\textbf {\bibinfo {volume} {25}},\ \bibinfo {pages} {155601} (\bibinfo
  {year} {2013})}\BibitemShut {NoStop}%
\bibitem [{\citenamefont {Meng}\ \emph {et~al.}(2013)\citenamefont {Meng},
  \citenamefont {Hung},\ and\ \citenamefont {Lang}}]{meng2013}%
  \BibitemOpen
  \bibfield  {author} {\bibinfo {author} {\bibfnamefont {Z.~Y.}\ \bibnamefont
  {Meng}}, \bibinfo {author} {\bibfnamefont {H.-H.}\ \bibnamefont {Hung}}, \
  and\ \bibinfo {author} {\bibfnamefont {T.~C.}\ \bibnamefont {Lang}},\
  }\href@noop {} {\bibfield  {journal} {\bibinfo  {journal} {Mod. Phys. Lett.
  B}\ }\textbf {\bibinfo {volume} {28}},\ \bibinfo {pages} {143001} (\bibinfo
  {year} {2013})}\BibitemShut {NoStop}%
\bibitem [{\citenamefont {Avron}\ \emph {et~al.}(1983)\citenamefont {Avron},
  \citenamefont {Seiler},\ and\ \citenamefont {Simon}}]{avron1983}%
  \BibitemOpen
  \bibfield  {author} {\bibinfo {author} {\bibfnamefont {J.~E.}\ \bibnamefont
  {Avron}}, \bibinfo {author} {\bibfnamefont {R.}~\bibnamefont {Seiler}}, \
  and\ \bibinfo {author} {\bibfnamefont {B.}~\bibnamefont {Simon}},\ }\href
  {\doibase 10.1103/PhysRevLett.51.51} {\bibfield  {journal} {\bibinfo
  {journal} {Phys. Rev. Lett.}\ }\textbf {\bibinfo {volume} {51}},\ \bibinfo
  {pages} {51} (\bibinfo {year} {1983})}\BibitemShut {NoStop}%
\bibitem [{\citenamefont {Kane}\ and\ \citenamefont
  {Mele}(2005{\natexlab{b}})}]{kane2005b}%
  \BibitemOpen
  \bibfield  {author} {\bibinfo {author} {\bibfnamefont {C.~L.}\ \bibnamefont
  {Kane}}\ and\ \bibinfo {author} {\bibfnamefont {E.~J.}\ \bibnamefont
  {Mele}},\ }\href {\doibase 10.1103/PhysRevLett.95.146802} {\bibfield
  {journal} {\bibinfo  {journal} {Phys. Rev. Lett.}\ }\textbf {\bibinfo
  {volume} {95}},\ \bibinfo {pages} {146802} (\bibinfo {year}
  {2005}{\natexlab{b}})}\BibitemShut {NoStop}%
\bibitem [{\citenamefont {Maier}\ \emph {et~al.}(2005)\citenamefont {Maier},
  \citenamefont {Jarrell}, \citenamefont {Pruschke},\ and\ \citenamefont
  {Hettler}}]{Maier2005}%
  \BibitemOpen
  \bibfield  {author} {\bibinfo {author} {\bibfnamefont {T.}~\bibnamefont
  {Maier}}, \bibinfo {author} {\bibfnamefont {M.}~\bibnamefont {Jarrell}},
  \bibinfo {author} {\bibfnamefont {T.}~\bibnamefont {Pruschke}}, \ and\
  \bibinfo {author} {\bibfnamefont {M.~H.}\ \bibnamefont {Hettler}},\ }\href
  {\doibase 10.1103/RevModPhys.77.1027} {\bibfield  {journal} {\bibinfo
  {journal} {Rev. Mod. Phys.}\ }\textbf {\bibinfo {volume} {77}},\ \bibinfo
  {pages} {1027} (\bibinfo {year} {2005})}\BibitemShut {NoStop}%
\bibitem [{\citenamefont {Kotliar}\ \emph {et~al.}(2006)\citenamefont
  {Kotliar}, \citenamefont {Savrasov}, \citenamefont {Haule}, \citenamefont
  {Oudovenko}, \citenamefont {Parcollet},\ and\ \citenamefont
  {Marianetti}}]{Kotliar2006}%
  \BibitemOpen
  \bibfield  {author} {\bibinfo {author} {\bibfnamefont {G.}~\bibnamefont
  {Kotliar}}, \bibinfo {author} {\bibfnamefont {S.~Y.}\ \bibnamefont
  {Savrasov}}, \bibinfo {author} {\bibfnamefont {K.}~\bibnamefont {Haule}},
  \bibinfo {author} {\bibfnamefont {V.~S.}\ \bibnamefont {Oudovenko}}, \bibinfo
  {author} {\bibfnamefont {O.}~\bibnamefont {Parcollet}}, \ and\ \bibinfo
  {author} {\bibfnamefont {C.~A.}\ \bibnamefont {Marianetti}},\ }\href
  {\doibase 10.1103/RevModPhys.78.865} {\bibfield  {journal} {\bibinfo
  {journal} {Rev. Mod. Phys.}\ }\textbf {\bibinfo {volume} {78}},\ \bibinfo
  {pages} {865} (\bibinfo {year} {2006})}\BibitemShut {NoStop}%
\bibitem [{\citenamefont {Wu}\ \emph {et~al.}(2010)\citenamefont {Wu},
  \citenamefont {Chen}, \citenamefont {Tao}, \citenamefont {Tong},\ and\
  \citenamefont {Liu}}]{Weiwu2010}%
  \BibitemOpen
  \bibfield  {author} {\bibinfo {author} {\bibfnamefont {W.}~\bibnamefont
  {Wu}}, \bibinfo {author} {\bibfnamefont {Y.-H.}\ \bibnamefont {Chen}},
  \bibinfo {author} {\bibfnamefont {H.-S.}\ \bibnamefont {Tao}}, \bibinfo
  {author} {\bibfnamefont {N.-H.}\ \bibnamefont {Tong}}, \ and\ \bibinfo
  {author} {\bibfnamefont {W.-M.}\ \bibnamefont {Liu}},\ }\href {\doibase
  10.1103/PhysRevB.82.245102} {\bibfield  {journal} {\bibinfo  {journal} {Phys.
  Rev. B}\ }\textbf {\bibinfo {volume} {82}},\ \bibinfo {pages} {245102}
  (\bibinfo {year} {2010})}\BibitemShut {NoStop}%
\bibitem [{\citenamefont {Budich}\ \emph {et~al.}(2013)\citenamefont {Budich},
  \citenamefont {Trauzettel},\ and\ \citenamefont {Sangiovanni}}]{Budich2013}%
  \BibitemOpen
  \bibfield  {author} {\bibinfo {author} {\bibfnamefont {J.~C.}\ \bibnamefont
  {Budich}}, \bibinfo {author} {\bibfnamefont {B.}~\bibnamefont {Trauzettel}},
  \ and\ \bibinfo {author} {\bibfnamefont {G.}~\bibnamefont {Sangiovanni}},\
  }\href {\doibase 10.1103/PhysRevB.87.235104} {\bibfield  {journal} {\bibinfo
  {journal} {Phys. Rev. B}\ }\textbf {\bibinfo {volume} {87}},\ \bibinfo
  {pages} {235104} (\bibinfo {year} {2013})}\BibitemShut {NoStop}%
\bibitem [{\citenamefont {Ara\'ujo}\ \emph {et~al.}(2013)\citenamefont
  {Ara\'ujo}, \citenamefont {Castro},\ and\ \citenamefont
  {Sacramento}}]{Miguel2013}%
  \BibitemOpen
  \bibfield  {author} {\bibinfo {author} {\bibfnamefont {M.~A.~N.}\
  \bibnamefont {Ara\'ujo}}, \bibinfo {author} {\bibfnamefont {E.~V.}\
  \bibnamefont {Castro}}, \ and\ \bibinfo {author} {\bibfnamefont {P.~D.}\
  \bibnamefont {Sacramento}},\ }\href {\doibase 10.1103/PhysRevB.87.085109}
  {\bibfield  {journal} {\bibinfo  {journal} {Phys. Rev. B}\ }\textbf {\bibinfo
  {volume} {87}},\ \bibinfo {pages} {085109} (\bibinfo {year}
  {2013})}\BibitemShut {NoStop}%
\bibitem [{\citenamefont {Florens}\ and\ \citenamefont
  {Georges}(2004)}]{Florens2004}%
  \BibitemOpen
  \bibfield  {author} {\bibinfo {author} {\bibfnamefont {S.}~\bibnamefont
  {Florens}}\ and\ \bibinfo {author} {\bibfnamefont {A.}~\bibnamefont
  {Georges}},\ }\href {\doibase 10.1103/PhysRevB.70.035114} {\bibfield
  {journal} {\bibinfo  {journal} {Phys. Rev. B}\ }\textbf {\bibinfo {volume}
  {70}},\ \bibinfo {pages} {035114} (\bibinfo {year} {2004})}\BibitemShut
  {NoStop}%
\end{thebibliography}%
\end{document}